\documentclass[journal]{IEEEtran}
\usepackage[table]{xcolor}
\usepackage{cite}
\usepackage{graphicx}
\usepackage{lipsum}
\usepackage{stfloats}
\usepackage{bm}
\usepackage{amsmath,amssymb}
\usepackage{amsthm}
\usepackage{pifont}
\usepackage{bbding}
\usepackage{booktabs}
\usepackage{ctable}
\usepackage{threeparttable}
\usepackage{footmisc}
\usepackage{framed} 
\usepackage{array}
\usepackage{multirow}
\usepackage{longtable}
\usepackage{rotating}
\usepackage{fancyhdr}
\usepackage{lastpage}
\usepackage{layout}
\usepackage{wrapfig}
\usepackage{xcolor}
\usepackage{tabularx}
\usepackage{tabu}
\usepackage{enumitem}
\usepackage{url}
\newtheorem{myDef}{\textbf{Definition}}

\newtheorem{myPropos}{\textbf{Proposition}}
\graphicspath{{Figures/}}
\DeclareGraphicsExtensions{.png, .eps}

\begin{document}
	\title{Joint User Association and Bandwidth Allocation in Semantic Communication Networks}
	\author{Le Xia,~\IEEEmembership{Graduate Student~Member,~IEEE},
				 Yao Sun,~\IEEEmembership{Senior~Member,~IEEE},
				 Dusit Niyato,~\IEEEmembership{Fellow,~IEEE},\\
				 Xiaoqian Li,~\IEEEmembership{Member,~IEEE},
				 and Muhammad Ali Imran,~\IEEEmembership{Fellow,~IEEE}
	\thanks{
	
	Copyright (c) 2015 IEEE. Personal use of this material is permitted. However, permission to use this material for any other purposes must be obtained from the IEEE by sending a request to pubs-permissions@ieee.org.
	
	 Preliminary results of this work have been presented in part at the IEEE International Conference on Computer Communications Workshops (INFOCOM WKSHPS), 2022~\cite{9797984}. (\textit{Corresponding author: Yao Sun}.)
	
	Le Xia, Yao Sun, and Muhammad Ali Imran are with the James Watt School of Engineering, University of Glasgow, Glasgow G12 8QQ, UK (e-mail: l.xia.2@research.gla.ac.uk; \{Yao.Sun, Muhammad.Imran\}@glasgow.ac.uk).
	
	Dusit Niyato is with the School of Computer Science and Engineering, Nanyang Technological University, Singapore 639798 (e-mail: dniyato@ntu.edu.sg).
	
	Xiaoqian Li is with the National Key Lab on Communications, University of Electronic Science and Technology of China, Chengdu 611731, China (e-mail: xqli@uestc.edu.cn).
}
}	
	\maketitle
	\begin{abstract}
	Semantic communication (SemCom) has recently been considered a promising solution to guarantee high resource utilization and transmission reliability for future wireless networks.
	Nevertheless, the unique demand for background knowledge matching makes it challenging to achieve efficient wireless resource management for multiple users in SemCom-enabled networks (SC-Nets).
	To this end, this paper investigates SemCom from a networking perspective, where two fundamental problems of user association (UA) and bandwidth allocation (BA) are systematically addressed in the SC-Net.
	First, considering varying knowledge matching states between mobile users and associated base stations, we identify two general SC-Net scenarios, namely perfect knowledge matching-based SC-Net and imperfect knowledge matching-based SC-Net.
	Afterward, for each SC-Net scenario, we describe its distinctive semantic channel model from the semantic information theory perspective, whereby a concept of bit-rate-to-message-rate transformation is developed along with a new semantics-level metric, namely system throughput in message (STM), to measure the overall network performance.
	In this way, we then formulate a joint STM-maximization problem of UA and BA for each SC-Net scenario, followed by a corresponding optimal solution proposed.
	Numerical results in both scenarios demonstrate significant superiority and reliability of our solutions in the STM performance compared with two benchmarks.
	
	\end{abstract}
	

	\IEEEpeerreviewmaketitle
	
	\section{Introduction}
	\IEEEPARstart{C}{urrent} wireless networks are witnessing tremendous traffic demands to accommodate the upcoming pervasive application intelligence alongside large-capacity and multi-modal content delivery services, including text, image, speech, and video~\cite{saad2019vision,shen2021holistic}.
	In response to the ever-increasing data rates and stringent requirements for low latency and high reliability, it is foreseeable that available communication resources like spectrum or energy will gradually become scarce.
	Combined with the almost insurmountable Shannon limit, these destined bottlenecks are, therefore, motivating us to hunt for bold changes in the new design of future networks, i.e., making a paradigm shift from bit-based traditional communication to context-based~\textit{semantic communication} (SemCom)~\cite{weaver1953recent,strinati20216g,bao2011towards,luo2022semantic, xie2021deep, xie2020lite, weng2021semantic,xia2023wiservr,basu2014preserving,carnap1952outline,liu2022indirect}.
	The concept of SemCom was first introduced by Weaver in his landmark paper~\cite{weaver1953recent}, which explicitly categorizes communication problems into three levels, including the technical problem at the bit level, the semantic problem at the semantic level, and the effectiveness problem at the information exchange level.
	Nowadays, the technical problem has been thoroughly investigated in the light of classical Shannon information theory~\cite{shannon1948mathematical}, while the evolution toward SemCom is just beginning to take shape, with the core focus of meaning delivery rather than traditional bit transmission.
    
    To be concrete, a transmitter in SemCom first leverages background knowledge relevant to source messages to filters out irrelevant content and refine semantic features that only require fewer bits for transmission, the process of which is called semantic encoding.
    Once the destination receiver has the required knowledge, the local semantic interpreters are capable of accurately restoring the original meaning from the received bits, even if there are intolerable bit errors at the syntactic level.
    This process is called semantic decoding.
    Consequently, efficient exchanges for the desired information with ultra-low semantic ambiguity can be achieved in SemCom under equivalent background knowledge between source and destination, while significantly alleviating the resource scarcity~\cite{strinati20216g,bao2011towards,luo2022semantic}.
	
	As a matter of fact, there have been several noteworthy related works paving ways for the development of SemCom.
	Powered by advanced natural language processing (NLP) algorithms, the authors in~\cite{xie2021deep} and~\cite{xie2020lite} developed a Transformer-based text sentence similarity metric to measure the semantic performance in end-to-end SemCom systems.
	In parallel, two speech distortion ration-related semantic metrics are employed in~\cite{weng2021semantic} for testing the speech signals received via SemCom.
	Moreover,~\cite{xia2023wiservr} sought the possibility of applying SemCom into wireless virtual reality video delivery to realize high-performance feature extraction and semantic recovery.
	Apart from these semantic-transceiver-design related works, some researches on information-theoretic characterization for SemCom is also of paramount importance.
	The authors in~\cite{bao2011towards} and~\cite{basu2014preserving} quantitatively measured semantic entropy by putting forward a semantic channel coding theorem, which is based on the logical probability of messages proposed by Carnap and Bar-Hillel in~\cite{carnap1952outline}.
	Besides,~\cite{liu2022indirect} recently studied the semantic rate-distortion function of information source on the basis of its intrinsic state and extrinsic observation in the memoryless source case.

	Given the above preliminary works on link-level SemCom, we believe that it is time to move forward to the upper layer, i.e., investigating wireless SemCom from a networking perspective.
	In this respect, our main task lies on seeking the optimal wireless resource management strategy in the~\textit{SemCom-enabled network} (SC-Net) to optimize its overall network performance in a semantics-aware manner.
	Especially considering the unique demand for background knowledge matching between multiple mobile users (MUs) and multi-tier base stations (BSs), efficient resource management should still be indispensable in the SC-Net, which can yield a host of benefits, such as ensuring high quality of SemCom services and strengthening bandwidth utilization.
	
	Considering the novel paradigm of SC-Net, we are encountering three fundamental networking challenges as follows:
	\begin{itemize}
		\item \emph{Challenge 1: How to mathematically construct a reasonable semantic channel model in view of the characteristics of SemCom?}
		Different from the traditional bit-based channel models, the first priority in the semantic channel model is to mathematically characterize semantic information delivered from a source to its destination~\cite{liu2022indirect}.
		In particular, mismatched background knowledge between the semantic encoder and decoder can cause a certain degree of semantic ambiguity as well as information distortion~\cite{bao2011towards}. 
		Hence, the first challenging problem is how to sketch a reasonable semantic channel model based on different knowledge matching degrees from a semantic information theory perspective.
		\item \emph{Challenge 2: How to define a proper metric to measure the SemCom-related network performance?}
		Since the meaning of delivered messages, rather than transmitted bits, becomes the sole focus of SemCom, traditional performance metrics based on Shannon's legacy, such as system throughput in bit, are no longer applicable to measure the network performance of SC-Net.
		Given the unique semantic channel model, how to define a proper SemCom-related metric should be another challenge.
		\item \emph{Challenge 3: How to determine an optimal resource management strategy to maximize the SemCom-related performance of SC-Net?} 
		In the cellular network architecture, user association (UA) and bandwidth allocation (BA) are two key mechanisms to realize resource management~\cite{xu2021survey}.
		When it comes to SC-Net, besides practical constraints like limited bandwidth resources and single-BS association, varying degrees of knowledge matching between MUs and BSs should also impose new stringent criteria on the UA and BA.
		Especially noting that the SemCom-related network performance is linked with the stochasticity of source information generation, how to efficiently devise a joint optimal UA and BA strategy forms the third challenge.
	\end{itemize}
	
	To the best of our knowledge, no paper has addressed all these challenges before.
	In this paper, we mainly investigate the resource management problem in the downlink of SC-Net.
	By taking into account the unique knowledge matching mechanism in SemCom, two different SC-Net scenarios are identified along with their respective joint optimization problems in terms of UA and BA.
	Correspondingly, two effective solutions are proposed to achieve the optimal semantics-level performance of SC-Net.
    In a nutshell, our main contributions are summarized as follows:
    \begin{itemize}
		\item
		We first identify and formally define two general SC-Net scenarios based on all possible knowledge matching states between MUs and BSs, namely perfect knowledge matching (PKM)-based SC-Net and imperfect knowledge matching (IKM)-based SC-Net.
		We then mathematically describe the distinctive semantic channel capacity model for the PKM-based SC-Net scenario from a semantic information-theoretical perspective.
		Taking this as the baseline case, the semantic channel model of IKM-based SC-Net is systematically constructed.
		The above addresses the aforementioned~\textit{Challenge 1}.
		\item
		Given the unique semantic channel models of SC-Net, we leverage a bit-rate-to-message-rate (B2M) transformation function to measure the message rate of each SemCom-enabled link, whereby a new metric, namely system throughput in message (STM), is effectively developed to accurately characterize the overall network performance at the semantic level.
		This corresponds to the aforementioned~\textit{Challenge 2}.
		Moreover, two joint STM-maximization problems of UA and BA are formulated for the two SC-Net scenarios, respectively.
		\item
		Resource management solutions are derived separately under the two scenarios.
		For the deterministic optimization problem in the PKM-based SC-Net, we directly employ a primal-dual decomposition method with a Lagrange-multiplier method to obtain the optimal UA and BA strategy.
		Notably, for the case of IKM-based SC-Net with a stochastic optimization problem, we particularly devise a two-stage solution to tackle with it.
		The first stage exploits a chance-constrained model to transform the primal stochastic problem into a deterministic one by introducing a given semantic confidence level, followed by the second stage solution using an interior-point method and a heuristic algorithm to finalize the joint optimal solution of UA and BA.
		 Hence,~\textit{Challenge 3} is also well addressed.
		\item
		Extensive simulations are conducted for both SC-Net scenarios to evaluate the performance of proposed solutions.
		Compared with two baselines, numerical results demonstrate significant superiority of our solutions in terms of STM performance.
		Moreover, the importance of adequate knowledge matching is also revealed, which can ensure low semantic ambiguity and high message rates in SemCom.
	\end{itemize}
    
    The remainder of this paper is organized as follows.
    The next section first presents the semantic channel models of both PKM-based and IKM-based SC-Nets.
    Then for the two different SC-Nets, Section III and IV formulate their joint UA and BA optimization problems and propose the corresponding solutions, respectively.
    Numerical results are demonstrated and discussed in Section V, followed by conclusions in Section VI.
    
    \section{Semantic Communication Model}
    \begin{figure*}[ht]
		\centering
		\includegraphics[width=0.8\textwidth]{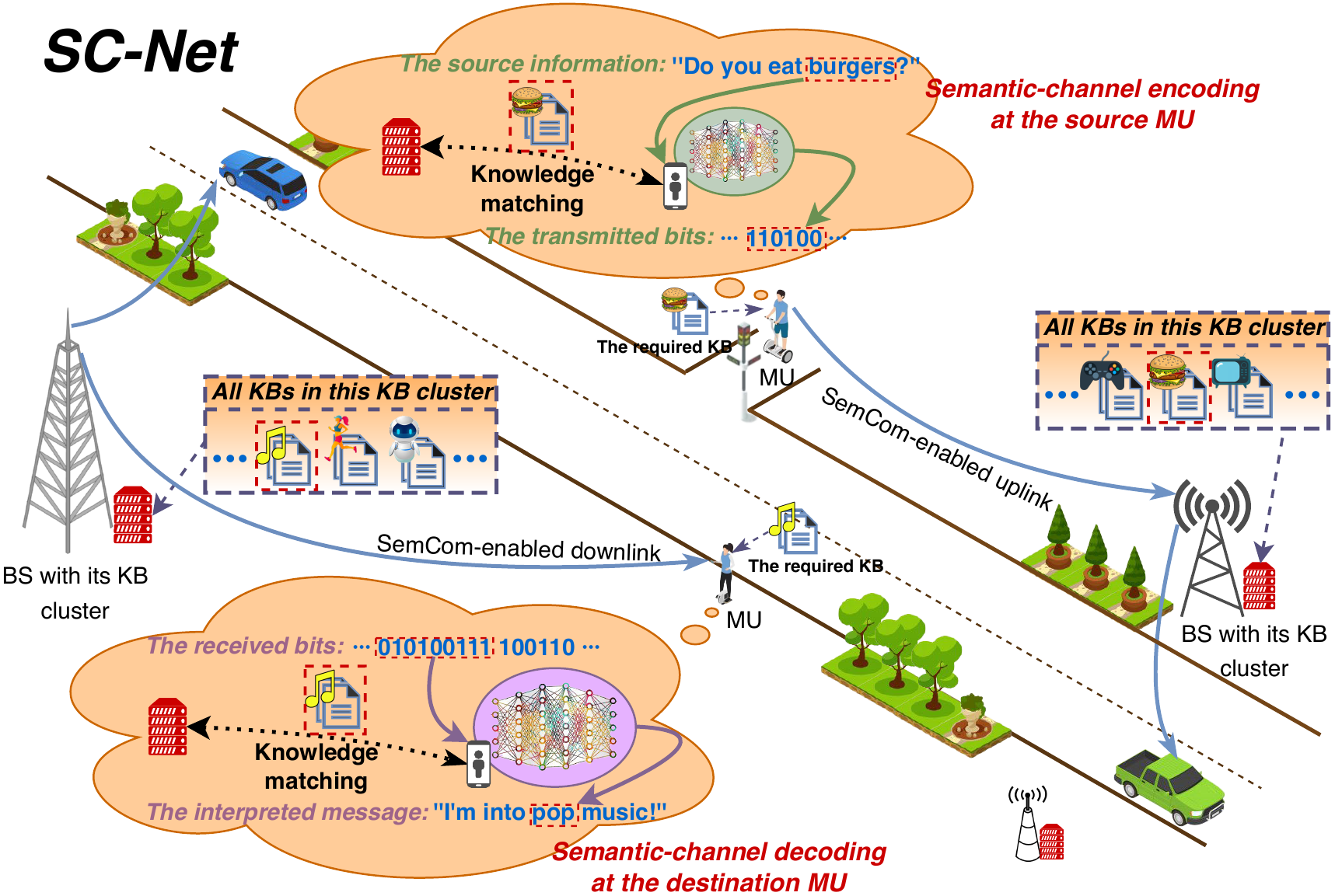} 
		\caption{An overview of SC-Net.}
		\label{SC-NetOverview}
    \end{figure*}
	\subsection{Background Knowledge Matching in SemCom}
	Consider an SC-Net scenario as shown in Fig.~\ref{SC-NetOverview}, where all communication parties (i.e., BSs and MUs) are capable of performing SemCom with each other.
	Recall that the accuracy of SemCom strongly relies on the matching degree of correct background knowledge between the transceiver (i.e., each pair of interrelated BS and MU), and the better knowledge matching degree is believed to guarantee lower semantic ambiguity and more efficient information interaction~\cite{xie2021deep,strinati20216g,bao2011towards}.
	Taking a single downlink in Fig.~\ref{SC-NetOverview} as an example, when a message related to the personal favorite music genre is delivered from a BS to an associated MU, they must have the same background knowledge in the musical domain so as to achieve accurate SemCom.
	In other words, the MU should ensure that its associated BS has the background knowledge that matches its own as closely as possible before requesting its desired SemCom services.
	
	On this basis, a key concept of auxiliary~\textit{knowledge base}\footnote{The structure of a KB can roughly cover multiple computational ontologies, facts, rules and constraints associated to a specific domain~\cite{chein2008graph}. In recent deep learning-driven semantic coding models, the KB is also treated as a training database serving a certain class of learning tasks~\cite{strinati20216g,luo2022semantic}. Relevant research is beyond the scope of this paper and will not be discussed in depth.} (KB) is introduced in SemCom, which is deemed a small information entity that stores the background knowledge of one particular application domain (such as music or sports) corresponding to a certain type of SemCom service~\cite{strinati20216g,shi2021semantic,luo2022semantic}.
	Combined with the powerful computation and storage ability of the BS, we further assume that each BS holds random amounts and types of KBs and name them a KB cluster, thus the MUs can acquire different SemCom services with required KBs by associating with different BSs.
	Nevertheless, it should be noted that messages received by each MU may cover differing background knowledge at the same time, leading to varying degrees of knowledge mismatch between the MU and its associated BS in the UA process.
	In this respect, we give our first definition as follows.
	\begin{figure}[!htbp]
		\centering
		\includegraphics[width=0.49\textwidth]{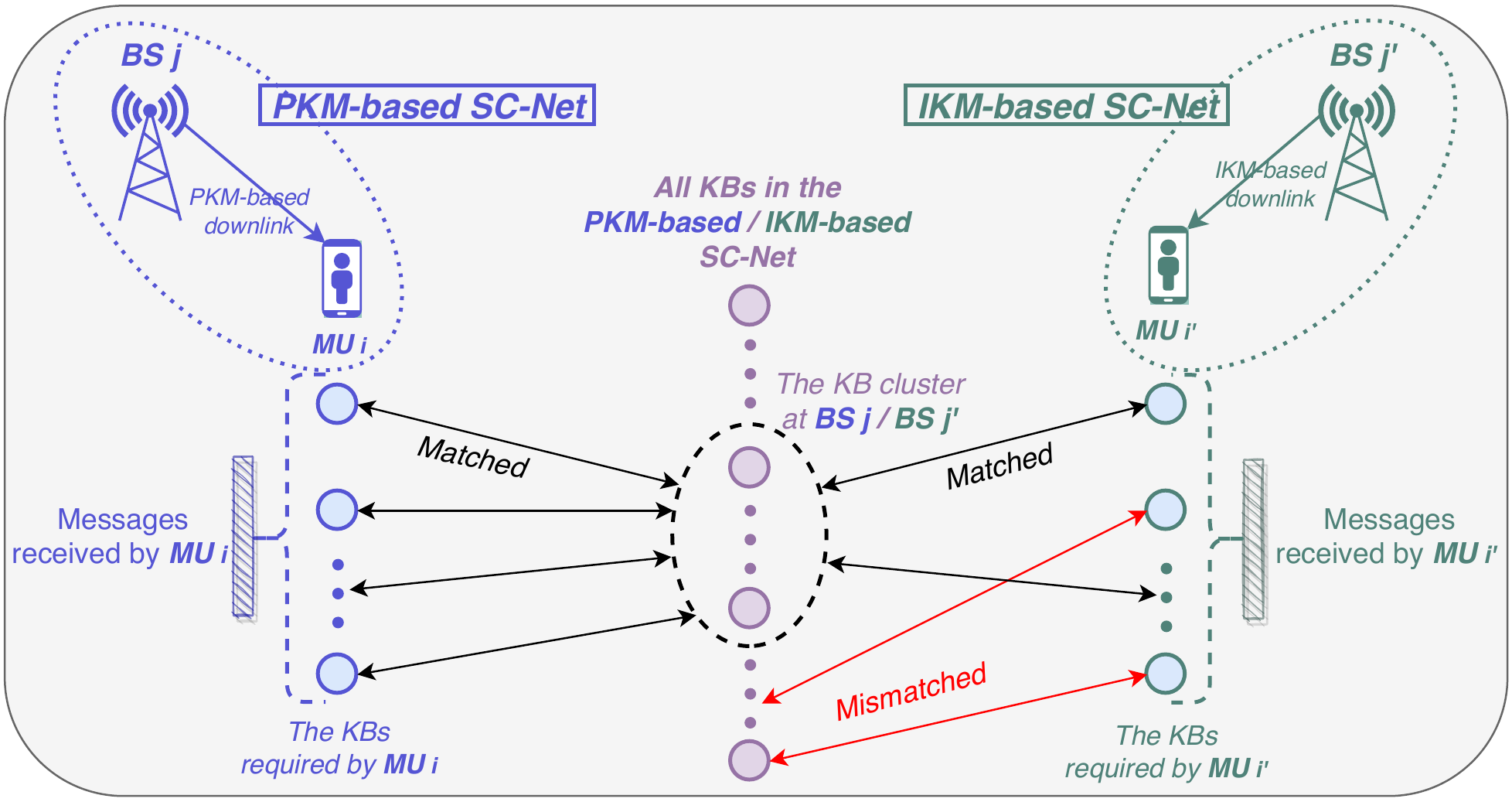} 
		\caption{Example illustration of the PKM-based SC-Net (on the left) and the IKM-based SC-Net (on the right) with respect to a single SemCom-enabled link.}
		\label{KBmatch}
    \end{figure}
	\begin{myDef}
		\textit{According to all possible knowledge matching cases, we define two different SC-Net scenarios, as illustrated in Fig.~\ref{KBmatch}.
		\begin{itemize}
			\item Perfect knowledge matching (PKM)-based SC-Net: For each MU in this network, there is at least one available BS holding all its required KBs to achieve perfect knowledge matching for SemCom.
			\item Imperfect knowledge matching (IKM)-based SC-Net: For each MU in this network, no BS holds all its required KBs, but its different associated BS may achieve varying degrees of (imperfect) knowledge matching for SemCom.
		\end{itemize}
		}
	\end{myDef}
	From Fig.~\ref{KBmatch}, it can be observed that MU $i$ on the left is categorized into the PKM case, as its associated BS $j$ precisely holds all KBs coherent with its received messages.
	As for MU $i'$ on the right, its associated BS $j'$ possesses only some of the required KBs, thereby only part of its received messages can be successfully interpreted, the case of which is defined as IKM.
	In the following two subsections, we will elaborate on the above two different SC-Net scenarios and their corresponding semantic channel models, respectively.
	
	\subsection{Semantic Channel Model in the PKM-based SC-Net}
\begin{figure}[!htbp]
		\centering
		\includegraphics[width=0.49\textwidth]{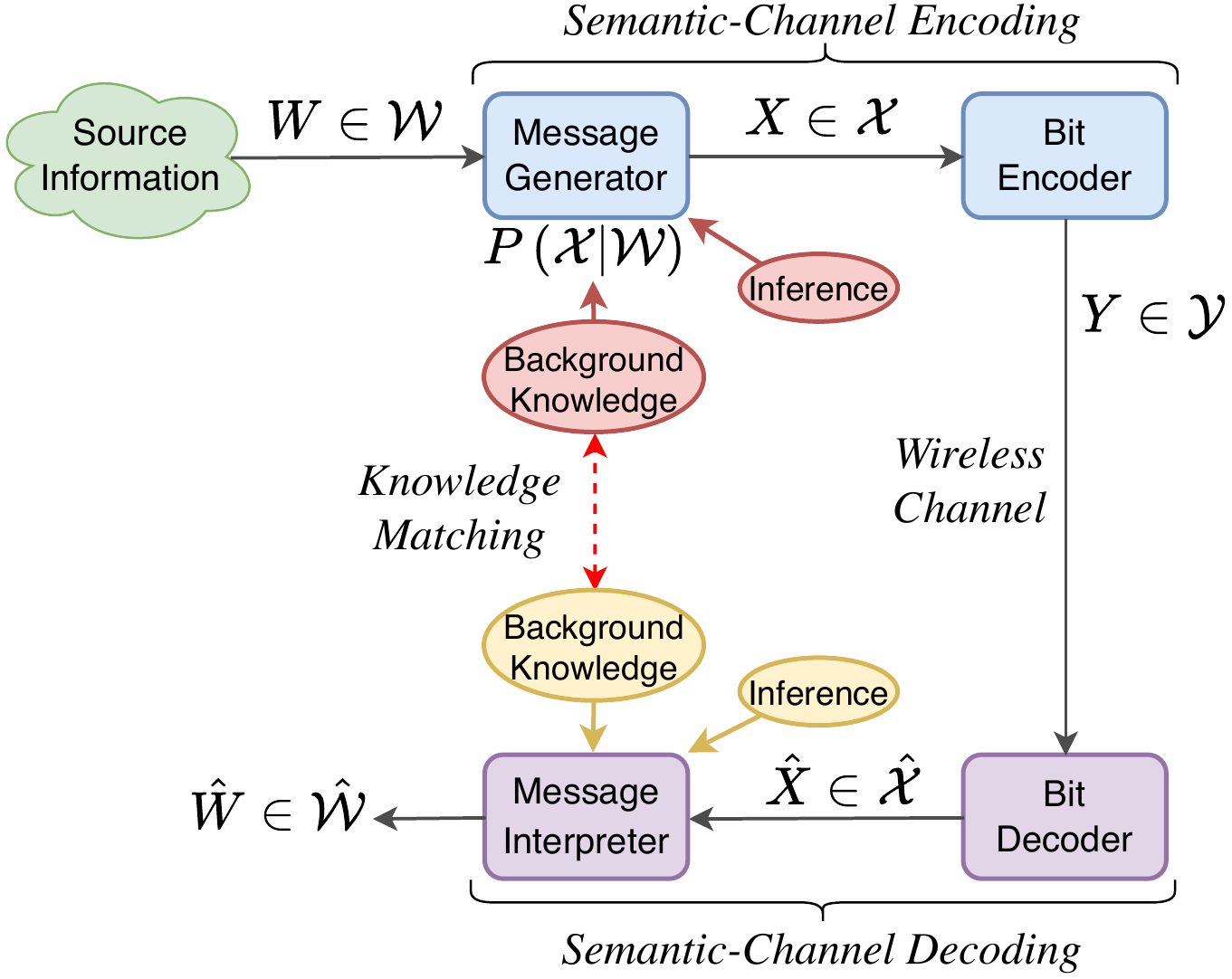} 
		\caption{A SemCom diagram of information source and destination.}
		\label{SemComChannel}
    \end{figure}

	Let us first consider a SemCom diagram as depicted in Fig.~\ref{SemComChannel}.
	Without loss of generality, the source information (i.e., the meaning desired to be conveyed) is modeled as a random variable $W$ and the generated observable message\footnote{The observable message here indicates a sequence of symbols syntactically (extrinsically) expressed in the language of the source, but actually contains specific information wished to be shared with the destination~\cite{bao2011towards}.} (e.g., a sentence or a speech signal representing the desired meaning) is denoted as $X$, which are defined over an appropriate product alphabet $\mathcal{W}\times\mathcal{X}$.
	Correspondingly, $\hat{X}\in \hat{\mathcal{X}}$ is the received message (e.g., the reconstructed sentence or speech), and $\hat{W} \in \hat{\mathcal{W}}$ is the interpreted information from $\hat{X}$ at the destination side.
	Among them, one bit encoder and one bit decoder are connected via a bit pipe (e.g., the wireless physical channel in traditional communications) to transmit the codeword $Y\in \mathcal{Y}$ at a certain code rate~\cite{liu2022indirect}.
	
	It is worth pointing out that the message generator of the source in Fig.~\ref{SemComChannel} is to generate $X$ from $W$ based on a specific semantic encoding strategy, and here we model the semantic encoding strategy as a conditional probabilistic distribution $P(\mathcal{X}| \mathcal{W})$ as that in~\cite{bao2011towards} and~\cite{liu2022indirect}.
	Meanwhile, we notice that different coding strategies $P(\mathcal{X}|\mathcal{W})$ can incur different degrees of semantic ambiguity from a statistical level~\cite{strinati20216g}, since a given observable message may semantically have more than one meaning while only some of them are true with respect to (w.r.t.) the source.\footnote{This is obvious and has been sufficiently demonstrated with examples in many existing studies~\cite{strinati20216g,bao2011towards,luo2022semantic}.}
	In order to guarantee adequate efficiency and accuracy for semantic coding, the background knowledge and the inference capability of each coding model become crucial in SemCom, as elucidated in~\cite{bao2011towards} and~\cite{basu2014preserving}.
	Herein, the inference capability can be understood as semantic coding models' feature compression and meaning interpretation abilities in the application case of deep learning-driven SemtCom, which are strongly correlated with the specific structures and composition of used neural networks~\cite{strinati20216g}.\footnote{For instance, in NLP-driven SemCom, attention-based models generally have a better inference capability than traditional recurrent (e.g., LSTM) or convolutional models (e.g., TextCNN) in the face of context prediction or sequence transduction related tasks~\cite{vaswani2017attention}.}
    
	Further proceeding as in~\cite{bao2011towards}, if the message generator and the message interpreter are assumed to have the identical inference capability and the perfectly matched background knowledge, the condition of Theorem 3 (semantic-channel coding theorem) proposed in~\cite{bao2011towards} can be fully met, stating that the semantic channel capacity (in units of messages per unit time,~\textit{msg/s}) of a discrete memoryless channel should be
	\begin{equation}
		C^{s}=\sup_{P(\mathcal{X}|\mathcal{W})}\left\{\sup_{P(\mathcal{Y}|\mathcal{X})}\left\{I(\mathcal{X};\hat{\mathcal{X}})\right\}-H(\mathcal{W}|\mathcal{X})+\overline{H_{s}(\hat{\mathcal{X}})}\right\}.\label{Cs}
	\end{equation}
	Here, $I(\mathcal{X};\hat{\mathcal{X}})$ is the mutual information between $\mathcal{X}$ and $\hat{\mathcal{X}}$ under the traditional bit encoding strategy (modeled as $P(\mathcal{Y}|\mathcal{X})$), while $H(\mathcal{W}|\mathcal{X})$ measures semantic ambiguity of coding at the source (w.r.t. $P(\mathcal{X}|\mathcal{W})$), both expressed in the form of classical Shannon entropy.
	Specially, unlike the above two terms, $\overline{H_{s}(\hat{\mathcal{X}})}$ measures the~\textit{semantic entropy} of received messages calculated by the~\textit{logical probability} (denoted as $P_{s}(\hat{X})$)~\cite{bao2011towards}, where $\overline{H_{s}(\hat{\mathcal{X}})}=-\sum_{\hat{X}\in \hat{\mathcal{X}}}P(\hat{X})\log_{2}P_{s}(\hat{X})$.\footnote{The concept of logical probability was first introduced by Carnap and Bar-Hillel in~\cite{carnap1952outline}, determining how likely it is for an observable message to be true, which is quite different from the common statistical probability $P(\hat{X})$. More technical details of $P_{s}(\hat{X})$ can refer to~\cite{strinati20216g,bao2011towards,basu2014preserving}, and~\cite{carnap1952outline}.}
	
	In this case, if denoting the optimal semantic encoding strategy as $P^{*}(\mathcal{X}|\mathcal{W})$, we can substitute it into~(\ref{Cs}) and obtain
	\begin{equation}
		\begin{aligned}
			C^{s}&=\sup_{P(\mathcal{Y}|\mathcal{X})} \left\{I^{*}(\mathcal{X};\hat{\mathcal{X}})\right\}-H^{*}(\mathcal{W}|\mathcal{X})+\overline{H_{s}^{*}(\hat{\mathcal{X}})}\\
			&\triangleq\sup_{P(\mathcal{Y}|\mathcal{X})} \left\{I^{*}(\mathcal{X};\hat{\mathcal{X}})\right\}+H^{s}\\
			& \triangleq C^{b}+H^{s},\label{CSP}
		\end{aligned}
	\end{equation}
	where $C^{b}$ characterizes the traditional Shannon channel capacity (in units of bits per unit time,~\textit{bit/s}), and $H^{s}=\overline{H_{s}^{*}(\hat{\mathcal{X}})}-H^{*}(\mathcal{W}|\mathcal{X})$ is a semantic-relevant term (can be positive or negative) depending on the background knowledge (i.e., the aforementioned KBs) and inference capability of specific semantic coding models adopted at the source and the destination.
	
	Keeping~(\ref{CSP}) in mind and let us now consider a single downlink channel between MU $i$ and BS $j$ in the PKM-based SC-Net.
	According to Definition 1, we first know that MU $i$ and BS $j$ must have the perfectly matched KBs.
	If further assuming BS $j$'s semantic encoder (i.e., message generator) has equal inference ability to MU $i$'s semantic decoder (i.e., message interpreter), it is seen that Theorem 3 in~\cite{bao2011towards} can be applied to this link.
	In line with~(\ref{CSP}), let $C^{s}_{ij}$ be its achievable message rate, let $C^{b}_{ij}$ be its achievable bit rate, and let $H^{s}_{ij}$ be its given semantic-relevant term of this link, thus we can formalize their relationship by giving the following definition.
	\begin{myDef}
		\textit{
		In the PKM-based SC-Net, we define $S^{P}_{ij}\left(\cdot \right)$ as the Bit-rate-to-Message-rate (B2M) transformation function of the physical link between MU $i$ and BS $j$, such that
		\begin{equation}
		S^{P}_{ij}\left(C^{b}_{ij}\right) \triangleq C^{s}_{ij} = C^{b}_{ij}+H^{s}_{ij}.
		\end{equation}
		}
	\end{myDef}
	
	In the light of Shannon theorem, we understand that $C^{b}_{ij}$ can be directly calculated based on the bandwidth and the signal-to-interference-plus-noise ratio (SINR) of the link.
	Hence, given the channel condition and the semantic coding models, we are able to adjust the bandwidth (i.e., the input of $S^{P}_{ij}\left(\cdot \right)$) allocated to this link so as to optimize the corresponding achievable message rate (i.e., the output of $S^{P}_{ij}\left(\cdot \right)$).
	Moreover, it can be observed that $S^{P}_{ij}\left(\cdot \right)$ is linear with bit rate $C^{b}_{ij}$, which renders a clear path towards the solution to the later PKM-based resource optimization problem.
	
	\subsection{Semantic Channel Model in the IKM-based SC-Net}
	When it comes to the IKM-based SC-Net, each communication link no longer satisfies the perfect knowledge matching condition so that the aforementioned semantic-channel coding theorem becomes inapplicable to any IKM-based case.
	More unfortunately, to the best of our knowledge, no work has proposed a SemCom-related information theory with rigorous derivations to declare the semantic channel capacity under mismatched background knowledge between the transceiver (i.e., the IKM case).
	Nevertheless, we note in this paper that for a given IKM-based link, there still exists an explicit relationship between the achievable message rate and the knowledge matching degree, i.e., the better the knowledge matching between source and destination, the more messages the destination can correctly interpret, and vice versa~\cite{strinati20216g}.
	The rationale behind this is quite speculative. For instance, we know from Definition 2 that the message rate is capable of reaching an upper bound $C^{s}_{ij}$ if MU $i$ and BS $j$ are in the PKM state, and conversely, no source information can be correctly interpreted if they have no matched KBs~\cite{bao2011towards}.
	
	In view of the above, the following definition describes our semantic channel modeling for the IKM case.
	\begin{myDef}
		\textit{In the IKM-based SC-Net, we define $S^{I}_{ij}\left(\cdot \right)$ as the B2M transformation function of the physical link between MU $i$ and BS $j$, which is correlated with its PKM-based B2M function $S^{P}_{ij}\left(\cdot \right)$ in the following manner
			\begin{equation}
				S^{I}_{ij}\left(\cdot \right)=\beta_{ij}\cdot S^{P}_{ij}\left(\cdot \right).\label{Def4}
			\end{equation}
		Here, $\beta_{ij}$ is named knowledge matching coefficient modeled as a random variable with the value ranging from $0$ to $1$, where $\beta_{ij}=0$ represents a completely mismatched state between MU $i$ and BS $j$, and $\beta_{ij}=1$ represents a perfectly matched state.}
	\end{myDef}
	
	Clearly, the upper bound of the message rate in the IKM case must be the message rate obtained in its PKM case (i.e., $\beta_{ij}=1$), while it is also able to reach zero when there is no common background knowledge between the transceiver (i.e., $\beta_{ij}=0$), as mentioned earlier.
	More importantly, since the information source is generally modeled as a stochastic process~\cite{liu2022indirect}, the specific amount of its generated messages corresponding to the matched KBs or the mismatched KBs becomes uncertain, even given the knowledge matching state.
	As a result, compared to the PKM case, there is always only a random proportion $\beta_{ij}$ of messages that can be correctly interpreted in the IKM case, eventually rendering a random message rate w.r.t. $S^{I}_{ij}\left(\cdot \right)$.
	In accordance with the above, we further make the following proposition.
	\begin{myPropos}
		Given the knowledge matching degree, denoted as $\tau_{ij}$, between MU $i$ and BS $j$ in the IKM case, the random knowledge matching coefficient $\beta_{ij}$ obeys a Gaussian distribution with mean $\tau_{ij}$ and variance $\sigma^{2}_{ij}$, i.e., $\beta_{ij}\sim \mathcal{N}\left(\tau_{ij},\sigma^{2}_{ij}\right)$, where $\sigma^{2}_{ij}=\tau_{ij}\left(1-\tau_{ij}\right)$.
	\end{myPropos}
	\begin{IEEEproof}
		Consider a downlink case between MU $i$ and BS $j$, first suppose that there is a set of source information modeled as a stochastic process $\left\{W_{m}\right\}$ ($m=1,2,\ldots,M$) at the BS $j$ side, where each $W_{m}$ is independent of each other~\cite{liu2022indirect}.
		Besides, let $K_{m}$ denote the KB required by source information $W_{m}$ for SemCom, and let $\mathcal{K}_{ij}$ denote the set of KBs matched between MU $i$ and BS $j$.
		Hence, we can define a KB matching indicator for source information as follows:
		\begin{equation}
				Z_{m}=\left\{\begin{aligned}
				1,\quad &  \text{if}\ K_{m}\in \mathcal{K}_{ij}\\
				0,\quad &  \text{otherwise}
			\end{aligned}.
			\right.
		\end{equation}
		
		Further given the knowledge matching degree $\tau_{ij}$ of the link between MU $i$ and BS $j$, thus the probability of successful matching of $W_{m}$, i.e., the probability of $Z_{m}=1$, becomes $\tau_{ij}$.
		Moreover, based on the same link, its different source information $\left\{W_{m}\right\}$ should have the same probability of successful matching, which is absolutely irrelevant to the knowledge matching situations of other links.
		As such, $\left\{Z_{m}\right\}$ obeys the identical binomial distribution w.r.t. $\tau_{ij}$, such that
		\begin{equation}
		\label{sdfgg}
		\begin{cases}
			\Pr(Z_{m}=1)=\tau_{ij}\\
			\Pr(Z_{m}=0)=1-\tau_{ij}
		\end{cases},
		\end{equation}
		where $\Pr(\cdot)$ is the probability measure.
		
		With these, the random knowledge matching coefficient $\beta_{ij}$ can be now expressed as the mean of the sum of $\left\{Z_{m}\right\}$ from a statistical average point of view as $M$ approaches infinity, that is,
		 \begin{equation}
		 	\beta_{ij}=\lim_{M\rightarrow +\infty}\frac{1}{M}\sum_{m=1}^{M}Z_{m}.\label{fdghh}
		 \end{equation}
		 
		 Based on (\ref{sdfgg}) and (\ref{fdghh}), the classical central limit theorem~\cite{fischer2011history} can be directly applied to determine the distribution of $\beta_{ij}$, whereby we have $\beta_{ij}\sim \mathcal{N}\left(\tau_{ij},\tau_{ij}\left(1-\tau_{ij}\right)\right)$.
	\end{IEEEproof}
	
	From Proposition 1, it is observed that in the IKM-based SC-Net case, each BS $j$ is capable of only getting the deterministic information of $\tau_{ij}$ (i.e., the distribution of $\beta_{ij}$) from its link associated with MU $i$, which always leads to a stochastic optimization problem for IKM-based resource management. Therefore, the IKM problem is inevitable, and its solution should be quite distinct from that of the PKM problem due to the stochasticity of each $\beta_{ij}$.
	
	\subsection{Basic Network Topology of SC-Net}
	Let us now consider the network topology of both PKM-based and IKM-based SC-Nets.
	As shown in Fig.~\ref{SC-NetOverview}, suppose that there are a total of $U$ MUs randomly located within the coverage of $B$ BSs, in which each MU $i \in \mathcal{U}=\left\{1,2,\ldots,U\right\}$ can only be associated with one BS $j \in \mathcal{B}=\left\{1,2,\ldots,B\right\}$ at a time.
	Specially, in alignment with Definition 1, first let $\mathcal{B}_{i}^{P}$ ($\mathcal{B}_{i}^{P}\subseteq \mathcal{B}$, $\forall i \in \mathcal{U}$) denote the set of BSs holding all the KBs required by MU $i$.
	As for the case of IKM-based SC-Net, assuming there is a minimum threshold for the knowledge matching degree, denoted as $\tau_{0}$, to guarantee the minimum quality of SemCom.
	That way, let $\mathcal{B}_{i}^{I}$ denote the set of BSs that MU $i$ is eligible for (user association) UA in the IKM-based SC-Net, where $\mathcal{B}_{i}^{I}=\left\{j\mid j \in\mathcal{B},\tau_{ij}\geqslant\tau_{0}\right\}$, $\forall i \in \mathcal{U}$.
	Based on the above, if we define the binary UA indicator for both scenarios as $x_{ij}\in \left\{ 0,1\right\}$, where $x_{ij}=1$ means that MU $i$ is associated with BS $j$ and $x_{ij}=0$ otherwise, the UA constraints for MU $i$ in the PKM-based and IKM-based SC-Nets are defined as follows:
	\begin{equation}
		\sum_{j\in \mathcal{B}_{i}^{P}} x_{ij}= 1 \ \ \text{and}\ \sum_{j\in \mathcal{B}_{i}^{I}} x_{ij}= 1,\ \ \forall i \in \mathcal{U},\label{UAconstraint}
	\end{equation}
	respectively.

	In the meantime, the total budget for bandwidth allocation (BA) of BS $j$ is denoted as $N_{j}$, and the amount of bandwidth that the BS $j$ assigns to MU $i$ is denoted as $n_{ij}$.
	Let $\gamma_{ij}$ be the SINR experienced by the link, so the achievable bit rate can be found by $C^{b}_{ij}=n_{ij}\log_{2}\left( 1+\gamma_{ij}\right)$.
	Further according to Definition 2 and Definition 3, the corresponding achievable message rate is $S^{P}_{ij}\left(C^{b}_{ij}\right)$ in the PKM-based SC-Net and $S^{I}_{ij}\left(C^{b}_{ij}\right)$ in the IKM-based SC-Net.
	With these, considering the uniqueness and significance of message rate in SemCom (i.e., the conveyed message itself becomes the sole focus of correct reception in SemCom rather than traditional transmitted bits~\cite{bao2011towards,strinati20216g}), we define a new performance metric herein, namely~\textit{system throughput in message} (STM), to specifically measure the overall message rates obtained by all MUs in the network.
	Consequently, the STM of PKM-based SC-Net is given as
	\begin{equation}
		\begin{aligned}
			\mathit{STM}^{P}&=\sum_{i\in \mathcal{U}}\sum_{j\in \mathcal{B}}x_{ij}S^{P}_{ij}\left(C^{b}_{ij}\right)\\
			&=\sum_{i\in \mathcal{U}}\sum_{j\in \mathcal{B}}x_{ij}S^{P}_{ij}\left(n_{ij}\log_{2}\left( 1+\gamma_{ij}\right)\right).
		\end{aligned}
	\end{equation}
	Likewise, the STM of IKM-based SC-Net is
	\begin{equation}
		\begin{aligned}
			\mathit{STM}^{I}&=\sum_{i\in \mathcal{U}}\sum_{j\in \mathcal{B}}x_{ij}S^{I}_{ij}\left(C^{b}_{ij}\right)\\
			&=\sum_{i\in \mathcal{U}}\sum_{j\in \mathcal{B}}x_{ij}\beta_{ij}S^{P}_{ij}\left(n_{ij}\log_{2}\left( 1+\gamma_{ij}\right)\right).
		\end{aligned}
	\end{equation}
	Based on $\mathit{STM}^{P}$ and $\mathit{STM}^{I}$, we are now able to jointly optimize the UA and BA from the SemCom perspective so as to respectively maximize the overall network performance for the two SC-Net scenarios.
	
	\section{Resource Management in the PKM-based SC-Net}
	\subsection{Problem Formulation}
	In order to empower very high quality of SemCom services for all MUs in the PKM-based SC-Net, it is of paramount importance to achieve the optimality of $\mathit{STM}^{P}$ subject to several SemCom-related and practical system constraints.
	To that end, we formulate an STM-maximization problem in a joint optimization manner of the UA variable $x_{ij}$ and the BA variable $n_{ij}$.
	For ease of illustration, hereafter we define a matrix $\bm{x}=\left\{x_{ij}\mid i\in\mathcal{U},j\in\mathcal{B}\right\}$ and a matrix $\bm{n}=\left\{n_{ij}\mid i\in\mathcal{U},j\in\mathcal{B}\right\}$ consisting of all variables related UA and BA, respectively.
	Note that both $\bm{x}$ and $\bm{n}$ are strongly correlated to semantic components, for example, $\bm{x}$ is strictly constrained by PKM-based links and $\bm{n}$ determines the upper bound of not only the bit-based channel capacity but also the semantic channel capacity.
	To be specific, the joint optimization problem of PKM-based SC-Net is given as follows:
	\begin{align}
	\mathbf{P1}:\ \max_{\bm{x},\bm{n}} \quad & \sum_{i\in \mathcal{U}}\sum_{j\in \mathcal{B}}x_{ij}S^{P}_{ij}\left(n_{ij}\log_{2}\left( 1+\gamma_{ij}\right)\right)~\label{P1}\\
	{\rm s.t.} \quad & \sum_{j\in \mathcal{B}_{i}^{P}} x_{ij}= 1,\  \forall i\in \mathcal{U},\tag{\ref{P1}a}\\
	&\sum_{i\in \mathcal{U}}x_{ij} n_{ij}\leqslant N_{j},\ \forall j\in \mathcal{B},\tag{\ref{P1}b}\\
	& x_{ij}\in \left\{ 0,1\right\},\ \forall \left( i,j\right) \in \mathcal{U}\times \mathcal{B}.\tag{\ref{P1}c}
	\end{align}
	Constraint~(\ref{P1}a) refers to the aforementioned single-BS constraint for UA, which also ensures that only the BSs in $\mathcal{B}_{i}^{P}$ can associate with MU $i$ to achieve the PKM state.
	Constraint~(\ref{P1}b) represents that the total bandwidth allocated to MUs cannot exceed the BA budget of each BS, and constraint~(\ref{P1}c) characterizes the binary property of $\bm{x}$.
	
	\subsection{Optimal Solution for UA}	
	Since the main difficulty of solving~$\mathbf{P1}$ lies on the $0$-$1$ constraint in~(\ref{P1}c), we first relax $\bm{x}$ into the continuous variable between $0$ and $1$.
	Notably, although we can directly solve the relaxed problem after the slack to $\bm{x}$, the most nontrivial point on recovering the binary property of $\bm{x}$ with low performance compromise is still intractable.
	To avoid this obstacle, in our solution, we assume that there is a minimum bandwidth amount that BS $j$ should allocate to its associated MU $i$, denoted as $n_{ij}^{T}$, to guarantee a basic quality of signal under the given channel condition.
	As such, by fixing each $n_{ij}$ as $n_{ij}^{T}$, $\mathbf{P1}$ can be rephrased as
	\begin{align}
	\mathbf{P1.1}:\ \max_{\bm{x}} \quad & \sum_{i\in \mathcal{U}}\sum_{j\in \mathcal{B}}x_{ij}\xi_{ij}^{T}~\label{P1.1}\\
	{\rm s.t.} \quad & \sum_{j\in \mathcal{B}_{i}^{P}} x_{ij}= 1,\  \forall i\in \mathcal{U},\tag{\ref{P1.1}a}\\
	&\sum_{i\in \mathcal{U}}x_{ij} n_{ij}^{T}\leqslant N_{j},\ \forall j\in \mathcal{B},\tag{\ref{P1.1}b}\\
	& 0\leqslant x_{ij}\leqslant 1,\ \forall \left( i,j\right) \in \mathcal{U}\times \mathcal{B},\tag{\ref{P1.1}c}
	\end{align}
	where
	\begin{equation}
		\xi_{ij}^{T}\triangleq S_{ij}^{P}\left(n_{ij}^{T}\log_{2}\left( 1+\gamma_{ij}\right)\right).\label{02}
	\end{equation}
	Notably, $\xi_{ij}^{T}$ is deemed a constant in the objective function~(\ref{P1.1}), since $n_{ij}^{T}$, $\gamma_{ij}$, and $H_{ij}^{s}$ in the B2M function $S_{ij}^{P}(\cdot)$ are all constants for the given link between MU $i$ and BS $j$.
	
	In the context of $\mathbf{P1.1}$, we employ the Lagrange dual method~\cite{low1999optimization} to obtain its dual optimization problem herein.
	By associating a Lagrange multiplier $\bm{\mu}=\left\{\mu_{j}\mid j\in \mathcal{B}\right\}$, the inequality constraint~(\ref{P1.1}b) can be incorporated into~(\ref{P1.1}), thereby its Lagrange function should be
	\begin{equation}
		L\left( x,\mu \right)=\sum_{i\in \mathcal{U}}\sum_{j\in \mathcal{B}}x_{ij}\xi_{ij}^{T}+\sum_{j\in \mathcal{B}}\mu_{j}\left(N_{j}-\sum_{i\in \mathcal{U}}x_{ij} n_{ij}^{T}\right).\label{Lag1.1}
	\end{equation}
	Hence, the Lagrange dual problem of $\mathbf{P1.1}$ becomes
	\begin{align}
			\mathbf{D1.1}:\ \min_{\bm{\mu}} \quad & D\left( \bm{\mu}\right)=g_{\bm{x}}\left( \bm{\mu}\right)+\sum_{j\in \mathcal{B}}\mu_{j}N_{j}\label{DDD}\\
			{\rm s.t.}\quad & \mu_{j}\geqslant 0,\ \forall j \in \mathcal{B},\tag{\ref{DDD}a}
	\end{align}
	where we have
	\begin{equation}
		\label{Dual}
			\begin{aligned}
			g_{\bm{x}}\left( \bm{\mu}\right)\ & =\ \sup_{\bm{x}}\  \sum_{i\in \mathcal{U}}\sum_{j\in \mathcal{B}}x_{ij}\left(\xi_{ij}^{T}-\mu_{j}n_{ij}^{T}\right)\\
			{\rm s.t.}\  & \ \text{(\ref{P1.1}a)},\ \text{(\ref{P1.1}c)}.
		\end{aligned}
	\end{equation}
	It is worth pointing out that strong duality holds in such primal-dual transformation, since the objective function~(\ref{P1.1}) of $\mathbf{P1.1}$ is convex and all its constraints are linear and affine inequalities, thus satisfying the Slater's condition~\cite{boyd2004convex}.
	
	Given the initial dual variable $\bm{\mu}$, we can first determine the optimal $\bm{x}$ (denoted as $\bm{x}^{*}=\left\{x_{ij}^{*}\mid i\in\mathcal{U},j\in\mathcal{B}\right\}$), and then leverage a gradient descent method~\cite{boyd2004convex} in charge of updating $\bm{\mu}$ to solve~$\mathbf{D1.1}$ in an iterative fashion.
	Carefully examining~(\ref{Dual}), it is easily derived that based on the fixed $n_{ij}^{T}$, MU $i$ can be served by its optimal BS $j$ if and only if it satisfies the following condition
	\begin{equation}
		\label{alpha}
			x_{ij}^{*}=\left\{\begin{aligned}
			1,\quad &  \text{if $j=\arg \max_{j\in \mathcal{B}_{i}^{P}}\left(\xi_{ij}^{T}-\mu_{j}n_{ij}^{T}\right)$}\\
			0,\quad &  \text{otherwise}
		\end{aligned},\ \forall i\in \mathcal{U}.
		\right.
	\end{equation}
	
	After getting $\bm{x}^{*}$, the gradient w.r.t. $\bm{\mu}$ in the objective function $D\left( \bm{\mu}\right)$ are calculated and set as the gradient in each iteration, whereby $\mu_{j}$ ($\forall j\in \mathcal{B}$) is updated as
	\begin{equation}
		\mu_{j}\left(t+1\right)=\left[\mu_{j}\left(t\right)-\delta\left(t\right)\cdot\left(N_{j}-\sum_{i\in \mathcal{U}}x_{ij}\left(t\right)n_{ij}^{T}\right) \right]^{+}.\label{TTTTT}
	\end{equation}
	The operator $\left[\cdot\right]^{+}$ here is to output the maximum value between its argument and zero, ensuring that $\bm{\mu}$ must be non-negative as constrained in~(\ref{DDD}a).
	$\delta\left(t\right)$ is the stepsize in iteration $t$ and generally, convergence of the gradient descent method can be guaranteed with the proper stepsize~\cite{wang2016joint}.
	Finally, to further ensure that the BA constraint~(\ref{P1.1}b) is not violated, the total amount of bandwidth consumed at each BS $j$ needs to be checked based on the obtained $\bm{x}^{*}$.
	For each BS that violates~(\ref{P1.1}b), we choose to reallocate its associated MUs who are consuming the most bandwidth to other BSs according to~(\ref{alpha}), until meeting the bandwidth budget requirements of all BSs.
	In summary, by alternatively updating $\bm{x}$ and $\bm{\mu}$ until convergence, the UA problem can be well solved in the PKM-based SC-Net.
	
	\subsection{Optimization Solution for BA}
	Given the obtained UA solution $\bm{x}^{*}$ and the fixed bandwidth threshold $n_{ij}^{T}$, we can directly formulate the BA problem for each BS $j$ ($\forall j\in \mathcal{B}$) as follows:
	\begin{align}
		\mathbf{P1.2^{(j)}}:\ \max_{\bm{n}} \quad & \sum_{i\in \mathcal{U}_{j}^{P}}S_{ij}^{P}\left(n_{ij}\log_{2}\left( 1+\gamma_{ij}\right)\right)~\label{P1.2}\\
		{\rm s.t.} \quad & \sum_{i\in \mathcal{U}_{j}^{P}}n_{ij}=N_{j},\tag{\ref{P1.2}a}\\
		\quad & n_{ij}\geqslant n_{ij}^{T},\ \forall i \in \mathcal{U}_{j}^{P},\tag{\ref{P1.2}b}
	\end{align}
	where
	\begin{equation}
		\mathcal{U}_{j}^{P}\triangleq\left\{ i\mid x_{ij}^{*}=1\right\}.
	\end{equation}
	Here, $\mathcal{U}_{j}^{P}$ stands for the set of MUs associated with BS $j$ in the previous UA phase.
	Owing to the linear property of $S_{ij}^{P}(\cdot)$, it is seen that for each $\mathbf{P1.2^{(j)}}$, the objective function as well as all constraints are convex, thereby some efficient optimization toolboxes such as CVPXY~\cite{diamond2016cvxpy} can be applied to directly finalize the optimal BA solution of PKM-based SC-Net.
	
	\section{Resource Management in the IKM-based SC-Net}
	
	\subsection{Problem Formulation}
	Similar to the rationale behind $\mathbf{P1}$, in the IKM-based SC-Net, achieving the optimality of $\mathit{STM}^{I}$ is also necessary for optimizing the overall SemCom-related network performance.
	Based on the UA indicator $\bm{x}$ and the BA indicator $\bm{n}$,\footnote{In order to avoid unnecessary redundant notations, in the IKM-based SC-Net, we use the same notations (e.g., $\bm{x}$ and $\bm{n}$, etc.) as in the PKM-based SC-Net, which have exactly the same physical meaning.} the joint optimization problem of IKM-based SC-Net can be formulated as
	\begin{align}
	\mathbf{P2}:\ \max_{\bm{x},\bm{n}} \quad & \sum_{i\in \mathcal{U}}\sum_{j\in \mathcal{B}}x_{ij}\beta_{ij}S^{P}_{ij}\left(n_{ij}\log_{2}\left( 1+\gamma_{ij}\right)\right)~\label{P2}\\
	{\rm s.t.} \quad & \sum_{j\in \mathcal{B}_{i}^{I}} x_{ij}= 1,\  \forall i\in \mathcal{U},\tag{\ref{P2}a}\\
	&\sum_{i\in \mathcal{U}}x_{ij} n_{ij}\leqslant N_{j},\ \forall j\in \mathcal{B},\tag{\ref{P2}b}\\
	& x_{ij}\in \left\{ 0,1\right\},\ \forall \left( i,j\right) \in \mathcal{U}\times \mathcal{B}.\tag{\ref{P2}c}
	\end{align}
	Different from $\mathbf{P1}$, the UA constraint~(\ref{P2}a) in $\mathbf{P2}$ ensures that only the IKM-enabled BSs in $\mathcal{B}_{i}^{I}$ can associate with MU $i$, which is determined by the given minimum knowledge matching threshold $\tau_{0}$ in the network.
	Likewise, constraints~(\ref{P2}b) and~(\ref{P2}c) represent the bandwidth budget limitation of each BS and the binary nature of $\bm{x}$, respectively.
	
	It is worth noting that the introduction of the random knowledge matching coefficient $\beta_{ij}$ leads to the biggest distinction between $\mathbf{P1}$ and $\mathbf{P2}$, where $\mathbf{P1}$ is clearly a deterministic optimization problem and $\mathbf{P2}$ is a stochastic optimization problem.
	That is, the solution $(\bm{x},\bm{n})$ to $\mathbf{P1}$ directly determines the numerical value of $\mathit{STM}^{P}$, while these two variables in $\mathbf{P2}$ actually affect the probability density function (PDF) of $\mathit{STM}^{I}$ (w.r.t. $\beta_{ij}$).
	Hence, the main difficulty of solving~$\mathbf{P2}$ lies on how to cope with the stochasticity of $\beta_{ij}$.
	In the paper, we dedicatedly develop a two-stage method to determine the optimal $\bm{x}$ and $\bm{n}$.
	Specifically, the first stage is to convert the nondeterministic problem $\mathbf{P2}$ into a deterministic one by leveraging a chance-constrained optimization model.
	Afterward, we devise an effective heuristic algorithm in the second stage to finalize the solution of UA and BA for the IKM-based SC-Net.
	
	\subsection{Problem Transformation with Semantic Confidence Level}
	Carefully examining~$\mathbf{P2}$, it can be seen that the random variable $\bm{\beta}=\left\{\beta_{ij}\mid i\in \mathcal{U}, j\in \mathcal{B}\right\}$ only exists in its objective function~(\ref{P2}).
	By taking into account the distribution of~(\ref{P2}), in our first-stage solution, we employ Kataoka's model~\cite{kataoka1963stochastic} to introduce a new objective function along with an extra constraint to make the primal problem suitable for stochastic optimization without altering the original intention.
	Denoting the new objective function as $\bar{F}(\bm{x},\bm{n})$ (which expression will be given later), according to~\cite{kataoka1963stochastic}, $\mathbf{P2}$ can be equivalently transformed into
	\begin{align}
	\mathbf{P2.1}:\ \max_{\bm{x},\bm{n}} \quad & \bar{F}(\bm{x},\bm{n})~\label{P2.1}\\
	{\rm s.t.} \quad & \Pr\left\{\mathit{STM}^{I}\geqslant \bar{F}(\bm{x},\bm{n})\right\}\geqslant \alpha,\tag{\ref{P2.1}a}\\
	&\text{(\ref{P2}a}),\ \text{(\ref{P2}b}),\ \text{(\ref{P2}c}).\tag{\ref{P2.1}b}
	\end{align}
	Constraint~(\ref{P2.1}a) is the newly introduced probabilistic (chance) constraint by a prescribed confidence level $\alpha$ ($0<\alpha <1$, large in practice~\cite{charnes1959chance}).
	To be more explicit, due to the randomness of $\beta_{ij}$, the goal of $\mathbf{P2.1}$ becomes to reach the optimality of $\left(\bm{x},\bm{n}\right)$ to determine the optimal PDF of $\mathit{STM}^{I}$, whereby its lower bound $\bar{F}(\bm{x},\bm{n})$ can be maximized based on the given confidence level $\alpha$.
	In this case, we name $\alpha$ as a semantic confidence level preset for the IKM-based SC-Net.\footnote{An expected value of optimization goal seems to be also applicable for the measure of optimality criterion~\cite{charnes1959chance}. However, the dispersion of random variables' distribution leads to a greater risk of getting a very low profit under the given expectation, as explained in~\cite{kataoka1963stochastic}. Hence, we set a given probability instead of the expected value in order to seek a higher practicality.}
	
	Besides, it can be observed that in the case of the optimal solution to $\mathbf{P2.1}$, $\mathit{STM}^{I}$ has a nondegenerate distribution (i.e., it does not reduce to a constant~\cite{prekopa2013stochastic}), which means
	\begin{equation}
		\Pr\left\{\mathit{STM}^{I}\geqslant \bar{F}(\bm{x},\bm{n})\right\}=\alpha\label{Probaconstraint}
	\end{equation}
	should have the same bound effect as constraint~(\ref{P2.1}a) to reach the optimality of~$\mathbf{P2.1}$.
	According to our Proposition 1, the sufficient condition of Theorem 10.4.1 proposed in~\cite{prekopa2013stochastic} is fully satisfied, which is to determine the specific expression of $\bar{F}(\bm{x},\bm{n})$ from~(\ref{Probaconstraint}).
	As such, we obtain
	\begin{equation}
	\label{JJJJJJ}
		\begin{aligned}
			\bar{F}(\bm{x},&\bm{n})= \sum_{i\in \mathcal{U}}\sum_{j\in \mathcal{B}}x_{ij}\tau_{ij}S_{ij}^{P}\left(n_{ij}\log_{2}\left( 1+\gamma_{ij}\right)\right)\\
			& -\Phi^{-1}(\alpha)\sqrt{\sum_{i\in \mathcal{U}}\left(\sum_{j\in \mathcal{B}}x_{ij}\sigma_{ij}S_{ij}^{P}(n_{ij}\log_{2}\left( 1+\gamma_{ij}\right))\right)^{2}},
		\end{aligned}
	\end{equation}
	where $\Phi^{-1}(\cdot)$ is the inverse function of the standard normal probability distribution.
	In view of~(\ref{Probaconstraint}) and~(\ref{JJJJJJ}), we see that even the biggest value of $\bar{F}(\bm{x},\bm{n})$ (or in other words, all ($\bm{x},\bm{n}$)) satisfies the confidence constraint in~(\ref{P2.1}a).
	Therefore,~(\ref{P2.1}a) can now be eliminated in $\mathbf{P2.1}$.
	
	On this basis, here we adopt the same strategy as in~(\ref{02}) to make $\mathbf{P2.1}$ tractable, where each $n_{ij}$ in $\bm{n}$ is fixed by the given bandwidth threshold $n_{ij}^{T}$ in the IKM-based SC-Net.
	Meanwhile, the UA variable $\bm{x}$ is relaxed into continuous as well to deal with the NP-hard obstacle.
	Consequently, we can obtain a deterministic optimization problem as
		\begin{align}
		\mathbf{P2.2}:\ \max_{\bm{x}} \quad & \bar{F}(\bm{x})\triangleq\bar{F}(\bm{x},\bm{n})|_{n_{ij}=n_{ij}^{T}}~\label{P2.2}\\
		{\rm s.t.} \quad & \sum_{j\in \mathcal{B}_{i}^{I}} x_{ij}=1,\  \forall i\in \mathcal{U},\tag{\ref{P2.2}a}\\
		&\sum_{i\in \mathcal{U}}x_{ij} n_{ij}^{T}\leqslant N_{j},\ \forall j\in \mathcal{B},\tag{\ref{P2.2}b}\\
		& 0\leqslant x_{ij}\leqslant 1,\ \forall \left( i,j\right) \in \mathcal{U}\times \mathcal{B}.\tag{\ref{P2.2}c}
		\end{align}
	As declaimed in~\cite{kataoka1963stochastic} and~\cite{charnes1959chance}, the convexity of the objective function $\bar{F}(\bm{x})$ can be guaranteed if assuming $\alpha>1/2$, i.e., $\Phi^{-1}(\alpha)>0$.
	Such an assumption is quite reasonable and practical, since a too small $\alpha$ means a very high-level limit on solution space $(\bm{x},\bm{n})$ according to constraint~(\ref{P2.1}a), which may even cause the nonexistence of feasible solutions combined with other constraints in $\mathbf{P2.1}$.
	In addition, it should be noted that $n_{ij}^{T}$, $\tau_{ij}$, $\sigma_{ij}$, and $\alpha$ in $\mathbf{P2.2}$ should all be treated as known constants related to the link between MU $i$ and BS $j$ when solving this problem.
	
	\subsection{Solution Finalization for UA and BA}
	In our second-stage solution, we first utilize the interior-point method~\cite{potra2000interior}, to approximately formulate the inequality-constrained problem $\mathbf{P2.2}$ into an equality-constrained problem so as to efficiently approach the optimality.
	To be concrete, let $\varphi(\bm{x})$ be a logarithmic barrier associated with the BA constraint~(\ref{P2.2}b), where
	\begin{equation}
		\varphi(\bm{x})=\sum_{j\in \mathcal{B}}\log(N_{j}-\sum_{i\in \mathcal{U}}x_{ij} n_{ij}^{T}).
	\end{equation}
	That way, $\mathbf{P2.2}$ can be rephrased as 
	\begin{align}
	\mathbf{P2.3}:\ \max_{\bm{x}} \quad & \bar{F}(\bm{x})+r\cdot \varphi(\bm{x})\label{P2.3}\\
	{\rm s.t.} \quad & \text{(\ref{P2.2}a)},\ \text{(\ref{P2.2}c)}.\tag{\ref{P2.3}a}
	\end{align}
	Here, $r$ is a small positive scalar that sets the accuracy of the approximation, and as $r$ decreases to zero, the maximum of the new objective function as in~(\ref{P2.3}) is able to converge to the optimal solution to the primal problem~\cite{potra2000interior}.
	It is important to mention that (\ref{P2.3}) still holds the convexity since both $\bar{F}(\bm{x})$ and $\varphi(\bm{x})$ are convex.
	As such, we can easily find a set $\bm{x}(r)$ that contains all the optimal $x_{ij}$ w.r.t. a given $r$ for $\mathbf{P2.3}$.
	Furthermore, according to the sequential unconstrained minimization mechanism~\cite{fiacco1990nonlinear}, the optimal solution $\bm{x}$ to $\mathbf{P2.2}$ (denoted as $\hat{\bm{x}}$) can be eventually obtained by iteratively updating the descent value of $r$ until convergence.\footnote{The value of $r$ and its update rule will be well initialized at the beginning of the barrier method. More technical details can be found in~\cite{boyd2004convex}.}
	
	Nevertheless, such $\hat{\bm{x}}$ cannot guarantee the binary value for each $\hat{x}_{ij}$.
	Therefore, we devise a heuristic algorithm herein to finalize the optimal solution to $\mathbf{P2}$ (i.e., $\bm{x}^{*}$) based on the given $\hat{\bm{x}}$.
	Specifically, each $x_{ij}^{*}$ is determined according to the following rule
	\begin{equation}
		\label{projection}
			x_{ij}^{*}=\left\{\begin{aligned}
			1,\quad &  \text{if}\ j=\arg \max_{j\in \mathcal{B}_{i}^{I}}\hat{x}_{ij}\\
			0,\quad &  \text{otherwise}
		\end{aligned}
		,\  \forall i\in \mathcal{U}.\right.
	\end{equation}
	An implicit interpretation to~(\ref{projection}) is that each MU in the IKM state has multiple potentially associated BSs along with the corresponding optimal weights, i.e., $\hat{x}_{ij}$s, which are strongly correlated with the performance of STM.
	Therefore, each user can select the BS with the maximum weight for UA to pursue the highest overall network performance for the SC-Net.
	 
	 Nevertheless, the resulted bandwidth consumption may still exceed some BSs' budget after executing~(\ref{projection}).
	 In this regard, we utilize the same countermeasure as in the PKM case by reassigning these MUs who consume the most bandwidth of these BSs to other BSs, based on their weight list given in $\hat{\bm{x}}$, until the BA constraint~(\ref{P2.2}b) is satisfied at all BSs.
	 Afterward, similar to the rationale of solving $\mathbf{P1.2^{(j)}}$, we can further formulate the BA optimization problem for each BS $j$ ($j \in \mathcal{B}$) based on the obtained $\bm{x}^{*}$ and the fixed $n_{ij}^{T}$. That is,
	\begin{align}
		\mathbf{P2.4^{(j)}}:\ \max_{\bm{n}} \quad & \bar{F}(\bm{x}^{*},\bm{n})~\label{P2.4}\\
		{\rm s.t.} \quad &\sum_{i\in \mathcal{U}_{j}^{I}}n_{ij}=N_{j},\tag{\ref{P2.4}a}\\
		\quad & n_{ij}\geqslant n_{ij}^{T},\ \forall i \in \mathcal{U}_{j}^{I},\tag{\ref{P2.4}b}
	\end{align}
	where
	\begin{equation}
		\mathcal{U}_{j}^{I}=\left\{ i\mid i \in \mathcal{U}, x_{ij}^{*}=1\right\}.
	\end{equation}
	In each $\mathbf{P2.4^{(j)}}$, the objective function~(\ref{P2.4}) is clearly convex as $\bar{F}(\bm{x},\bm{n})$ is convex, and both constraints~(\ref{P2.4}a) and~(\ref{P2.4}b) are linear, to which the toolbox CVPXY can be applied as well~\cite{diamond2016cvxpy}. 
	Finally, both the UA and BA problems have been well optimized in the IKM-based SC-Net, even with the intervention of the random knowledge matching coefficient $\bm{\beta}$.
	
	\section{Numerical Results and Discussions}
	In this section, we evaluate the performance of our proposed UA and BA solutions for PKM-based and IKM-based SC-Nets, respectively.
	In the basic network settings, we randomly drop $5$ pico BSs (PBS), $10$ femto BSs (FBS), and $200$ MUs in a circular area with a radius of $500$ meters, where a macro BS (MBS) is placed at the circle center.
	Meanwhile, the transmit power of the MBS, PBSs, and FBSs is set to $43$ dBm, $35$ dBm, and $20$ dBm, respectively, each of which has a bandwidth budget of $2$ MHz.
	For the wireless propagation model, we use $L(d)=34+40log(d)$ and $L(d)=37+30log(d)$ as the path loss model of the MBS/PBSs and FBSs, respectively, while supposing there is a fixed noise power of $-111.45$ dBm~\cite{boostanimehr2014unified}.
	
	As for the SemCom-related model, we simulate a general text transmission-enabled SC-Net environment to examine the proposed solutions for accurate demonstration purposes.
	Note here that the transmission scenarios for other types of content (e.g., image or video) can also be simulated for performance test, and the reason we choose the text-based scenario is because that there already exist well-established NLP-driven SemCom models.
	To be specific, the Transformer with the same structure as proposed in~\cite{xie2021deep} is adopted as a unified semantic coding model for all SemCom-enabled links, and the PyTorch-based Adam optimizer is applied for network training with an initial learning rate of $1\times10^{-3}$.
	Apart from this, all the source information used for transmission is based on a public dataset from the proceedings of European Parliament~\cite{koehn2005europarl}, where all initial sentences are pruned into a given word-counting range from $4$ to $30$ to facilitate subsequent computing efficiency and avoid potential gradient vanishing or explosion.
	With these, the corresponding PKM-based B2M function $S_{ij}^{P}(\cdot)$ can be approximated from model testing and will be shown later in the results.
	In the solution simulation of the PKM-based case, we set a dynamic stepsize of $\delta(t)=0.8/t$ to update the Lagrange multipliers in~(\ref{TTTTT}), where the convergence of each trial can always be guaranteed.
	In the IKM-based case, the knowledge matching degree $\tau_{ij}$ (w.r.t. $\beta_{ij}$ in Proposition 1, $\forall \left( i,j\right) \in \mathcal{U}\times \mathcal{B}$) is unified to $0.5$ for all possible links in the SC-Net, and hereafter we omit the subscript $ij$ from $\tau_{ij}$ for expression brevity.
	Moreover, the semantic confidence level is set to $\alpha=95\%$ in the two-stage solution of the IKM case.
	
	For comparison purposes, we utilize two baselines of UA and BA algorithms for both the PKM-based and IKM-based SC-Nets: 1) A~\textit{max-SINR plus water-filling} algorithm~\cite{he2013water}, in which each MU is associated with the BS that can provide the strongest SINR in its UA phase with the water-filling BA method; 2) A~\textit{max-SINR plus evenly-distributed} algorithm~\cite{ye2013user} that adopts the same max-SINR strategy for UA and an evenly-distributed BA method.
	Furthermore, a bit rate threshold (w.r.t. $n_{ij}^{T}$) of $0.01$ Mbit/s is fixed in both the proposed and baseline solutions to ensure a basic quality of SemCom services for all MUs.
	Notably, all the above parameter values are set by default unless otherwise specified, and all subsequent simulation results are obtained by averaging over a significantly large number of trials.
		
	\subsection{Performance Evaluations in the PKM-based SC-Net}
\begin{figure*}[tp]
		\centering
		\begin{minipage}[t]{0.48\textwidth}
			\centering
			\includegraphics[width=8cm]{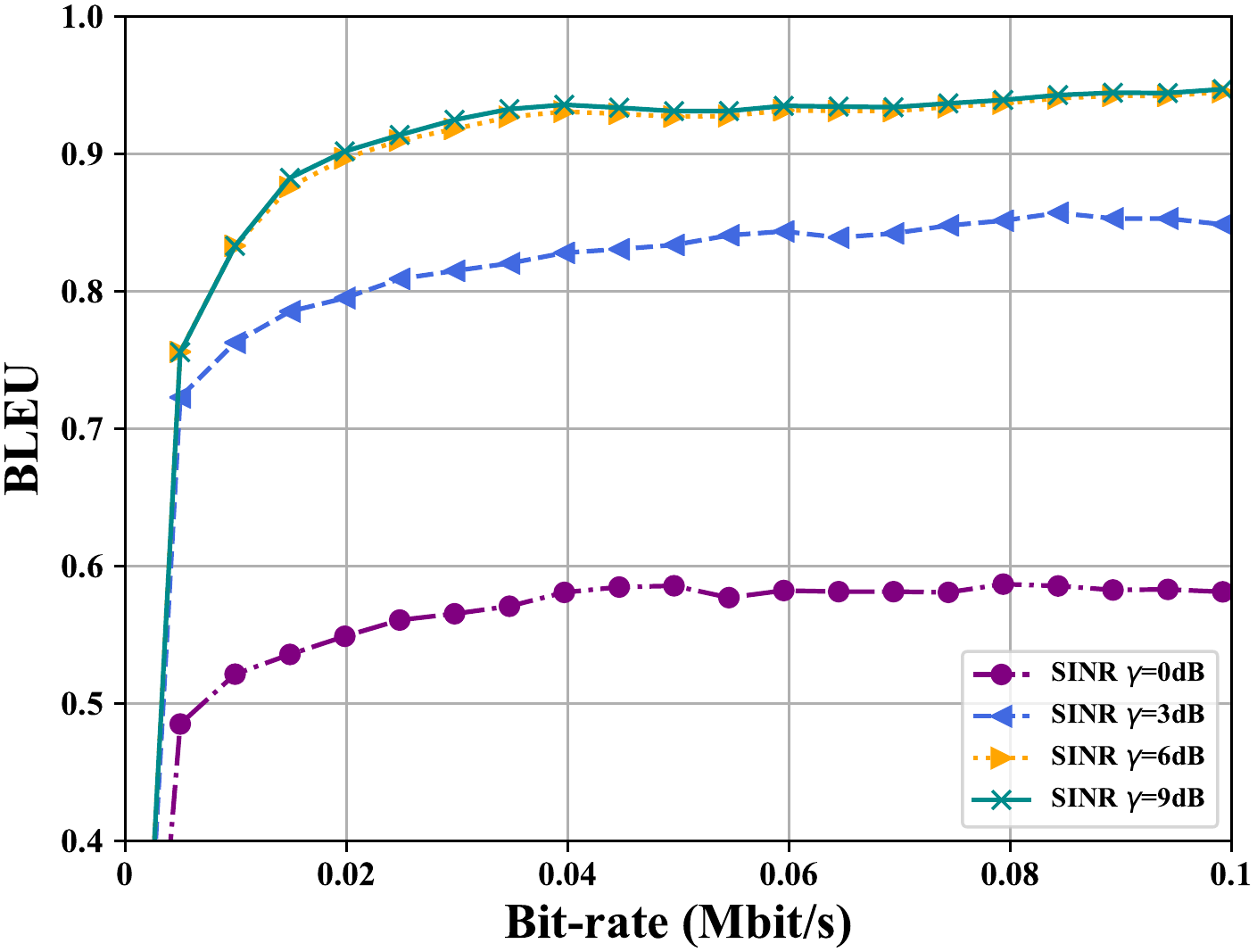}
			\caption{The BLEU score (1-gram) vs. bit rates under four different SINRs of $0$, $3$, $6$, and $9$ dB in the PKM-based SC-Net.}
			\label{BLEU}
		\end{minipage}
		\hspace{.15in}
		\begin{minipage}[t]{0.48\textwidth}
			\centering
			\includegraphics[width=8cm]{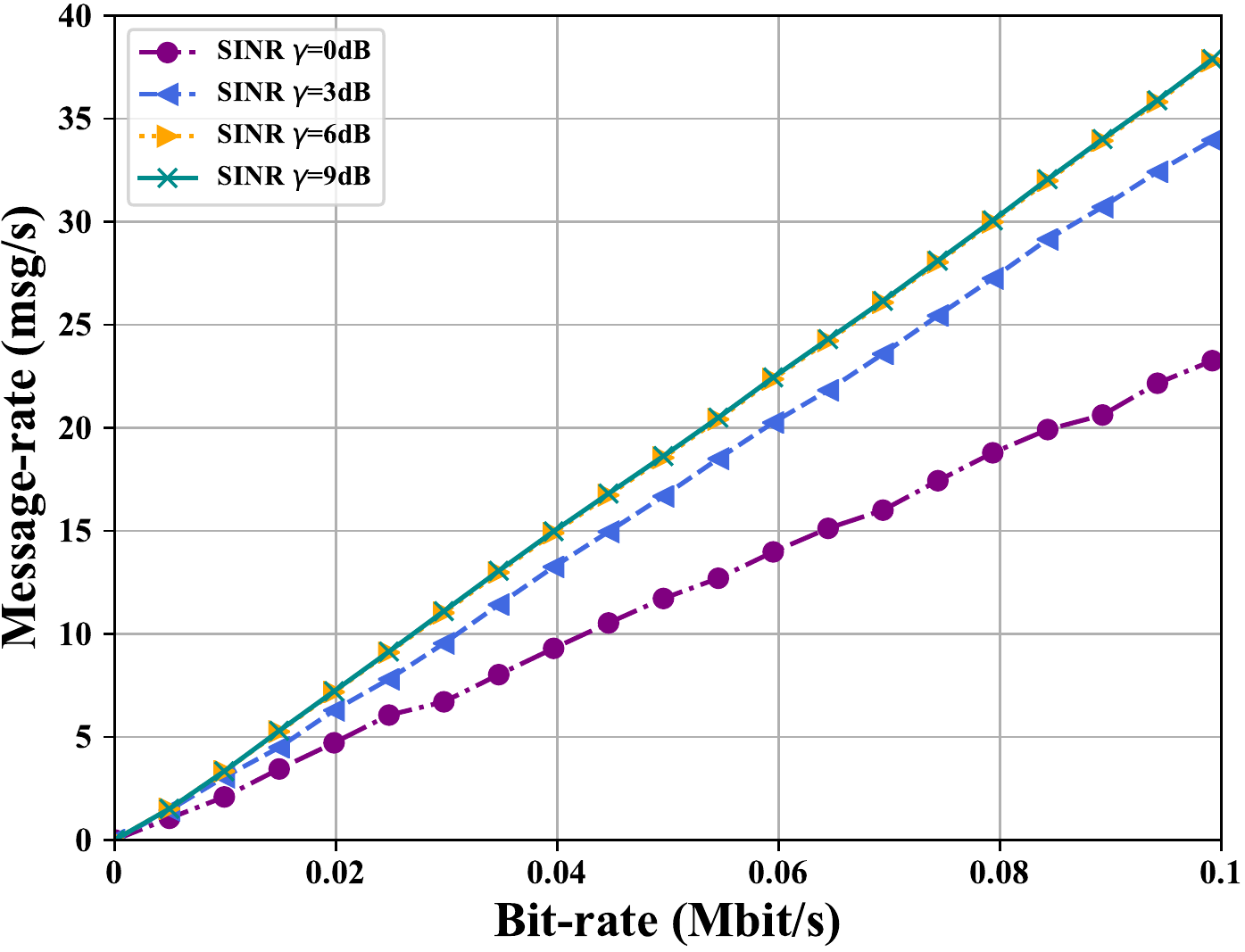}
			\caption{Demonstration of B2M transformation under four SINRs of Transformer-powered SemCom-enabled links in the PKM-based SC-Net.}
			\label{S_function}
		\end{minipage}
	\end{figure*}

	We first examine the performance of bilingual evaluation understudy (BLEU) in the PKM-based SC-Net, which is a classical metric in the NLP field with a value between $0$ and $1$~\cite{papineni2002bleu}.
	To be more concrete, it is scored via counting the word difference between the source and restored texts, and the closer its score is to 1, the better the text recovery.
	By testing the BLEU, the accuracy of semantic interpretation can be observed, which performance is also strongly related to the amount of messages MUs can correctly interpret in the network.
	As such, we first present the BLEU scores ($1$-gram) with different bit rates (i.e., $C_{ij}^{b}$) in Fig.~\ref{BLEU}, where four different SINRs are considered in the link.
	In this figure, it is seen that the BLEU under each SINR first grows as the bit rate improves, and soon stays at a stable score after about $0.03$ Mbit/s.
	Besides, we can observe a higher BLEU under a higher SINR, and when the SINR is larger than $6$ dB, the obtained BLEU scores are almost the same.
	This trend is predictable since the received bits can suffer different degrees of signal attenuation from different channel conditions, and obviously, the more correct bits the MU receives, the lower semantic ambiguity it achieves.
	Particularly, the above phenomena indicate a necessity of providing a minimum bit rate for MUs under good channel conditions in the SC-Net to achieve high-quality SemCom.
	
	\begin{figure*}[tp]
		\centering
		\begin{minipage}[t]{0.48\textwidth}
			\centering
			\includegraphics[width=8cm]{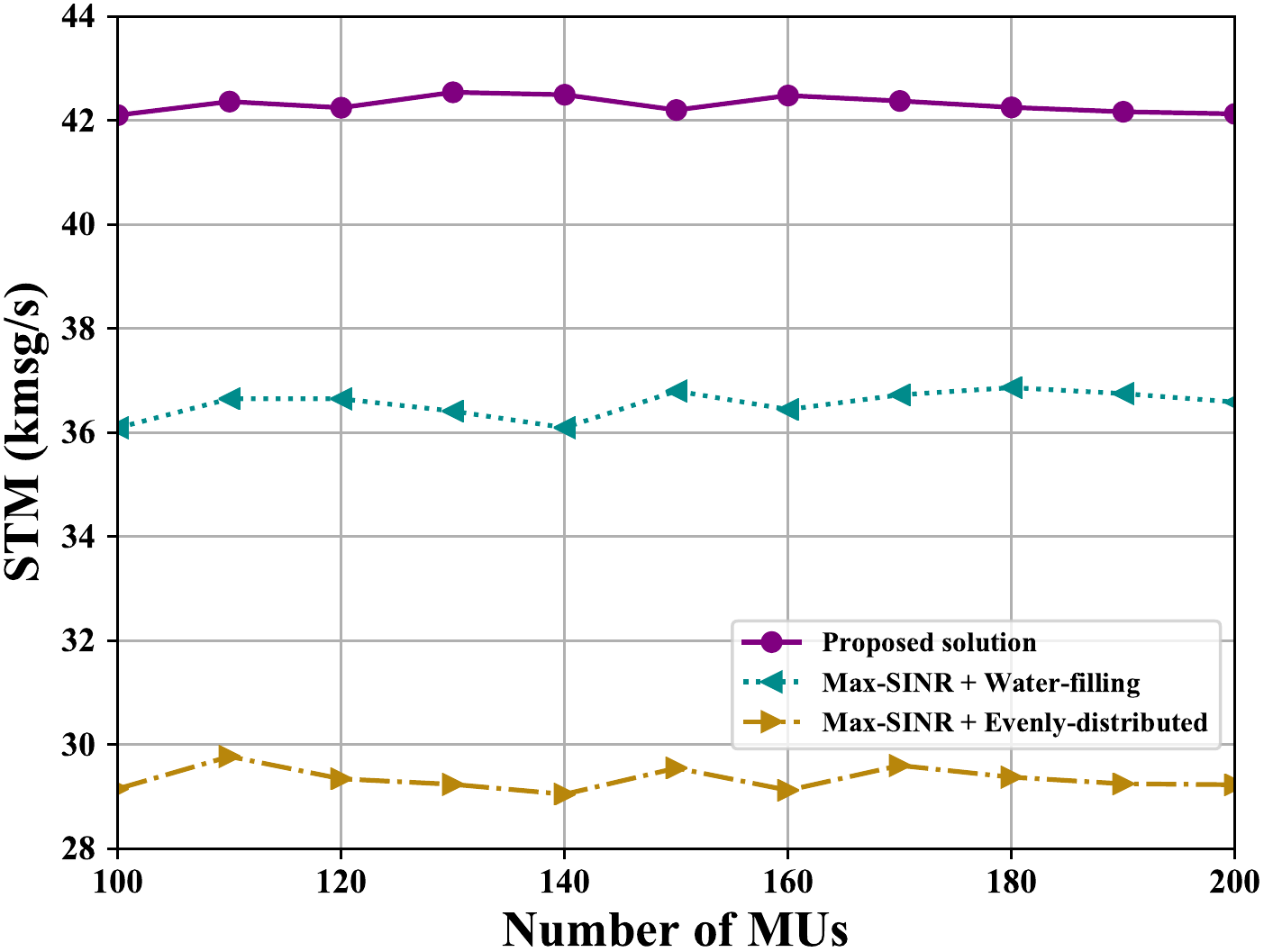}
			\caption{Comparison of the STM performance under different numbers of MUs in the PKM-based SC-Net.}
			\label{PKM_MU}
		\end{minipage}
		\hspace{.15in}
		\begin{minipage}[t]{0.48\textwidth}
			\centering
			\includegraphics[width=8cm]{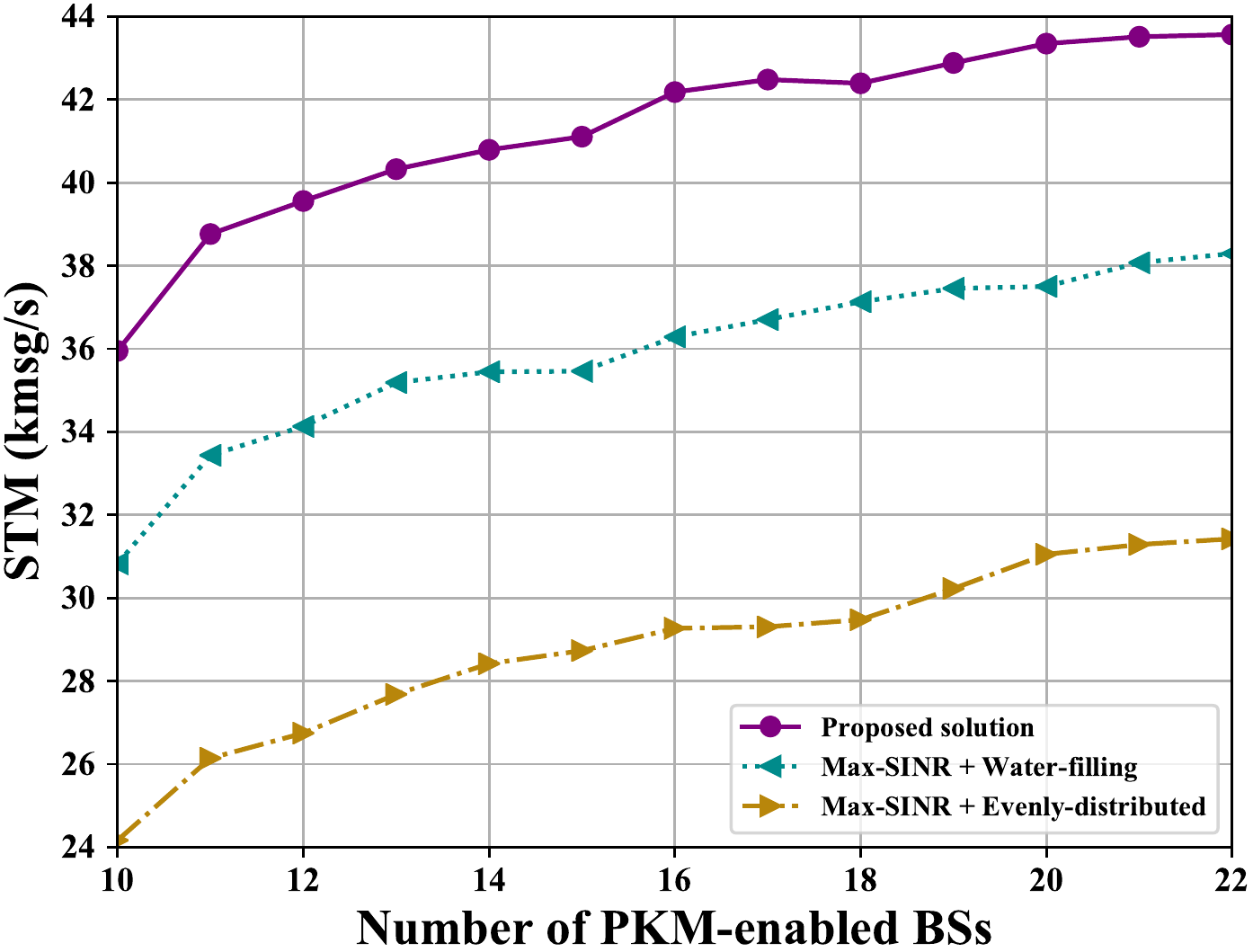}
			\caption{Comparison of the STM performance under different numbers of BSs in the PKM-based SC-Net.}
			\label{PKM_BS}
		\end{minipage}
	\end{figure*}
	
	According to~(\ref{CSP}), if the optimal semantic encoding strategy is guaranteed, it is believed that each MU can obtain a high message rate as the corresponding bit rate improves, which trend is related to its BLEU.
	To test this conjecture, we further draw the B2M transformation relationship (w.r.t. $S_{ij}^{P}(\cdot)$) under the same four SINRs in Fig.~\ref{S_function}.
	Notably, the message rate (i.e., $C_{ij}^{s}$) obtained here is based on calculating the amount of messages correctly interpreted in a given time unit.
	As expected, we can see that the transformed message rate grows at a steady rate with increasing the bit rate, and the better the SINR, the higher the transformation rate of B2M.
	
	The effectiveness of our UA and BA solution is demonstrated in the next two simulations, where the PKM-enabled BS means that each associated MU uses a well-trained Transformer decoding model under the perfectly matched training data.
	Fig.~\ref{PKM_MU} first compares the proposed solution with the two baselines by evaluating the STM performance under varying numbers of MUs between $100$ to $200$.
	It is seen that the STM obtained by our solution always far outperforms the two baselines.
	Specifically, the proposed solution always maintains an average STM at $42.5$ kmsg/s, which is around $6$ kmsg/s higher than the max-SINR plus water-filling baseline and $13$ kmsg/s than the max-SINR plus evenly-distributed baseline.
	Besides, the stable STM trend observed by all methods is because that the bandwidth budget of all BSs is reached in the BA phase, thus it is hard to improve the STM performance just by increasing the number of MUs.
	
	\begin{figure*}[tp]
		\centering
		\begin{minipage}[t]{0.48\textwidth}
			\centering
			\includegraphics[width=8cm]{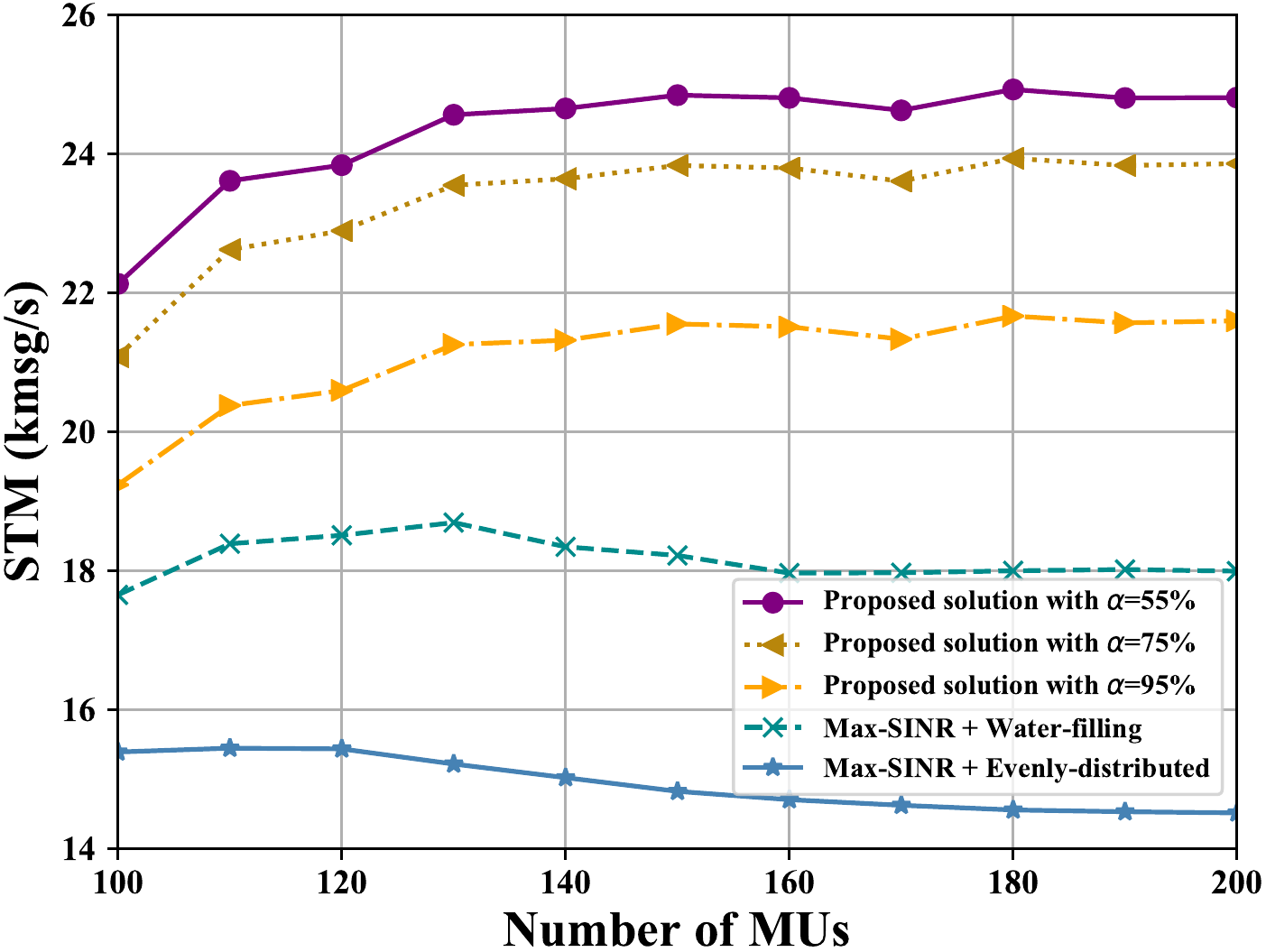}
			\caption{The STM performance against varying number of MUs under three semantic confidence levels in the IKM-based SC-Net.}
			\label{IKM_1}
		\end{minipage}
		\hspace{.15in}
		\begin{minipage}[t]{0.48\textwidth}
			\centering
			\includegraphics[width=8cm]{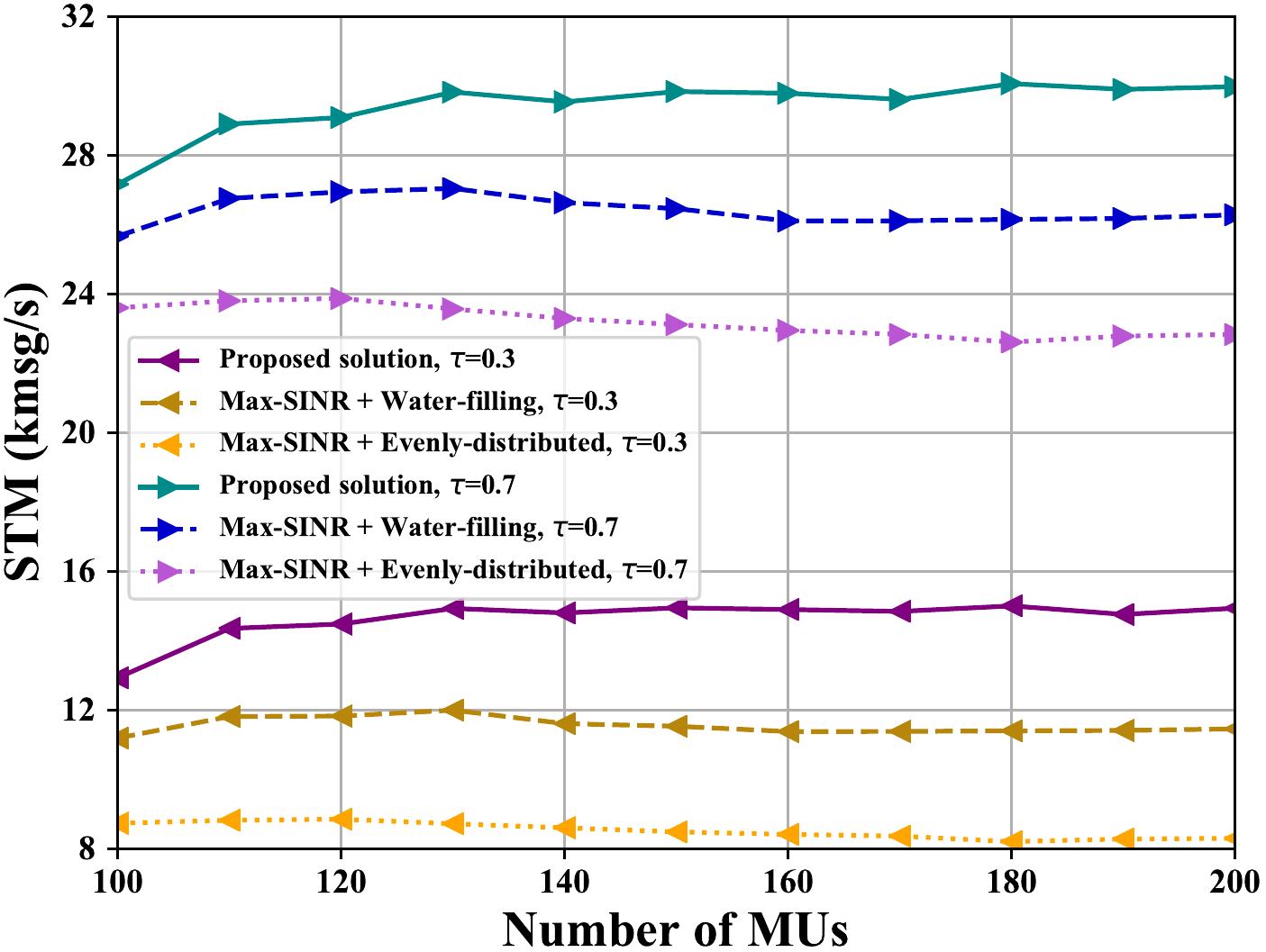}
			\caption{The STM performance against varying number of MUs under two average knowledge matching degrees in the IKM-based SC-Net.}
			\label{IKM_4}
		\end{minipage}
	\end{figure*}
	
	\begin{figure*}[tp]
		\centering
		\begin{minipage}[t]{0.48\textwidth}
			\centering
			\includegraphics[width=8cm]{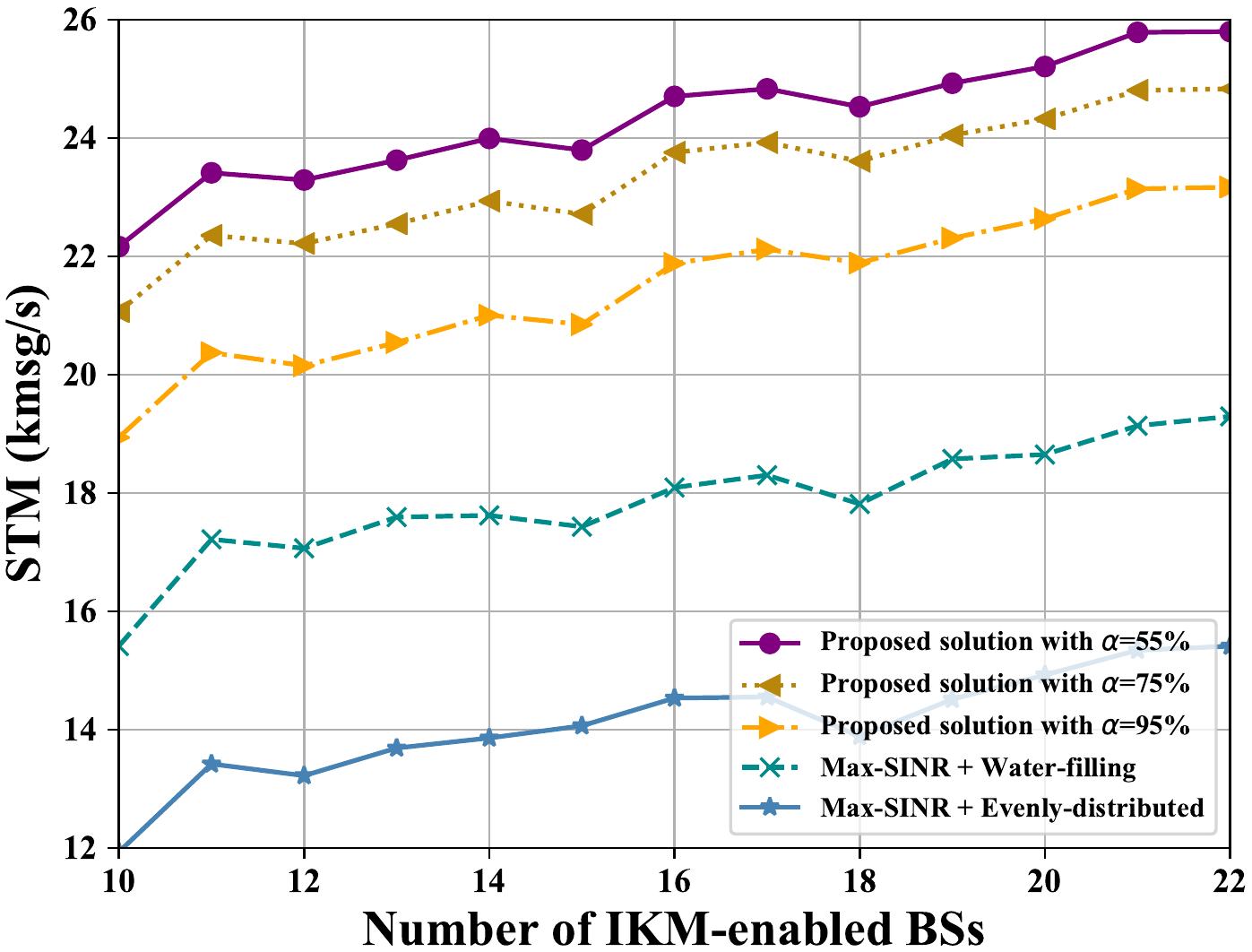}
			\caption{The STM performance against different numbers of BSs under three semantic confidence levels in the IKM-based SC-Net.}
			\label{IKM_3}
		\end{minipage}
		\hspace{.15in}
		\begin{minipage}[t]{0.48\textwidth}
			\centering
			\includegraphics[width=8cm]{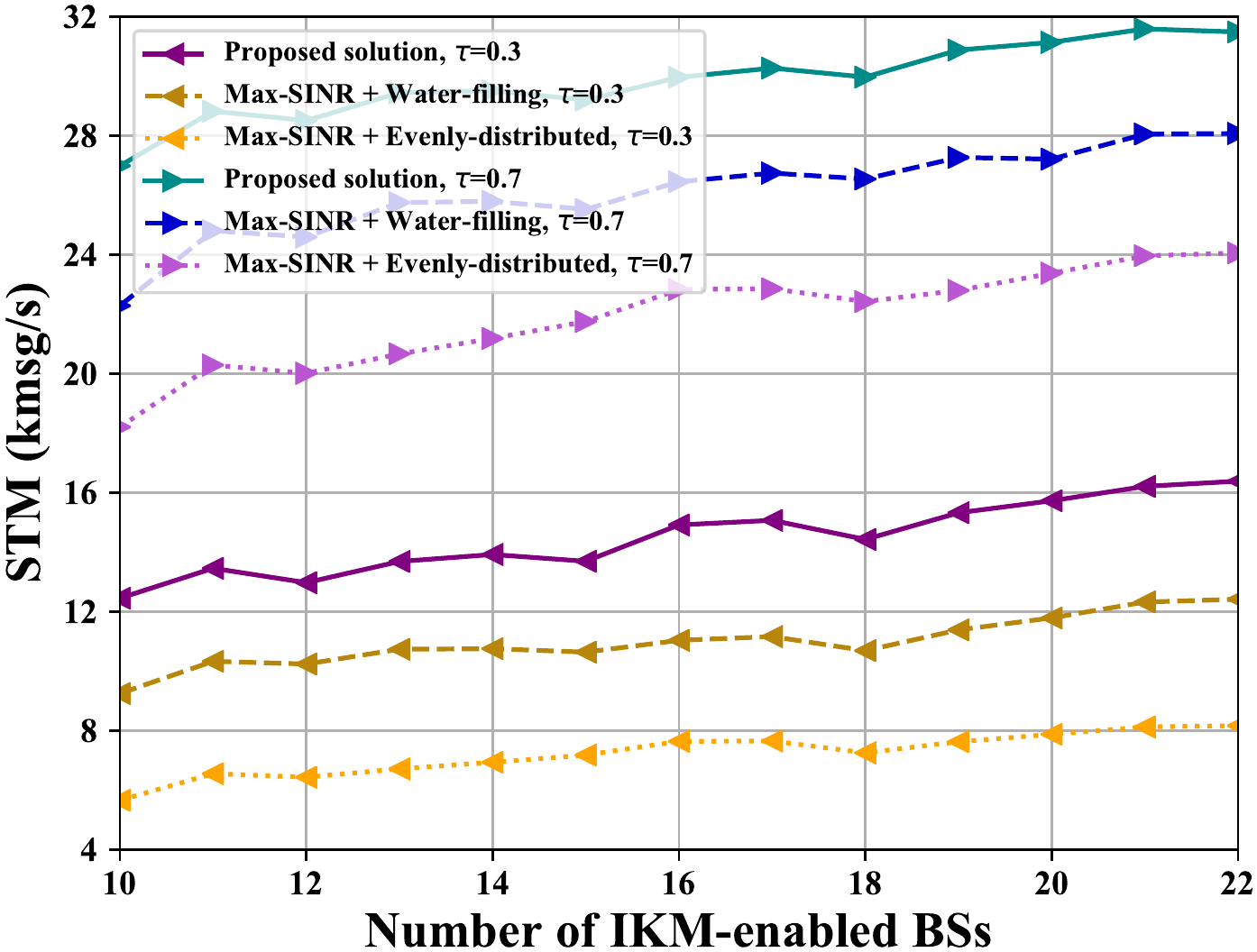}
			\caption{The STM performance against different numbers of BSs under two average knowledge matching degrees in the IKM-based SC-Net.}
			\label{IKM_2}
		\end{minipage}
	\end{figure*}
	
	Furthermore, similar comparisons are conducted with different numbers of PKM-enabled BSs between $10$ to $20$, as shown in Fig.~\ref{PKM_BS}.
	Consistent with the results in Fig.~\ref{PKM_MU}, the proposed solution gains an extra average STM around $6$ kmsg/s compared with the max-SINR plus water-filling baseline, and around $12$ kmsg/s compared with the max-SINR plus evenly-distributed baseline.
 	In the meantime, we can see that the STM performance of our solution increases at the beginning, and then gradually tends to stabilize after exceeding $20$ BSs.
 	As there are more BSs that can provide MUs with PKM-based SemCom services, the MUs will correspondingly have more bandwidth resources available to achieve higher message rates.
 	However, when the number of BSs surpasses a maximum threshold, the STM performance is believed to saturate and be even worsen, which is because of the severe channel interference incurred by the excess BSs.
	
	\subsection{Performance Evaluations in the IKM-based SC-Net}
	To evaluate the proposed two-stage solution in the IKM-based SC-Net, Fig.~\ref{IKM_1} first shows the comparisons of STM with different numbers of MUs, where three different semantic confidence levels of $\alpha=55\%$, $75\%$, and $95\%$ are taken into account.
	From this figure, we can see that the obtained STM increases with the number of MUs at the beginning, and soon remains stable after exceeding $130$ MUs.
	This is because the bandwidth budget of some BSs starts to be reached after serving the high number of MUs, thereby the STM is inevitably stabilized, as mentioned earlier.
	Moreover, an upward trend of STM performance is observed along with the reduction of $\alpha$, which can be explained from two perspectives.
	From the mathematical perspective, as introduced in~(\ref{Probaconstraint}), a higher value of $\alpha$ is equivalent to a lower achievable-bound on STM, which thus leads to the worse overall network performance.
	If we delve it in a semantical manner, due to the randomness of the knowledge matching degree in the IKM case, moderately increasing the preset semantic confidence can reduce the risk of getting below the expected message rate of each MU.
	Hence, some performance that compromises on STM are considered acceptable in alignment with the high preset semantic confidence level.
	Besides, it is seen in Fig.~\ref{IKM_1} that even at the highest required semantic confidence of $\alpha=95\%$, the proposed solution can consistently outperform the two benchmarks.
	
	The similar performance gain can be found in Fig.~\ref{IKM_4}, where each solution is performed under two mean knowledge matching degrees of $\tau=0.3$ and $0.7$ w.r.t. $\bm{\beta}$ as in Definition 3.
	To be explicit, under each $\tau$, our two-stage solution can always gain an extra STM performance around $2$ kmsg/s to $6$ kmsg/s with the increasing number of MUs when compared with the two baselines.
	In addition, the BA constraint of each method is always satisfied, hence, we can see the STM performance stabilizes from $130$ MUs, which trend is consistent with that in the previous figure.
	As for the impact of different knowledge matching degrees, it always shows a reducing trend of STM as $\tau$ decreases.
	Since the higher $\tau$ represents the larger likelihood of having a good knowledge matching for SemCom, each MU can correspondingly obtain a higher accuracy of message interpretation, so that a better STM performance renders in the IKM-based SC-Net.
    
%
    
    Next, we evaluate the STM performance with different semantic confidence levels $\alpha$ and knowledge matching degrees $\tau$ in Fig.~\ref{IKM_3} and Fig.~\ref{IKM_2}, respectively, under varying numbers of IKM-enabled BSs.
    Fig.~\ref{IKM_3} first presents the STM at different $\alpha$, and the higher STM is seen again by the lower semantic confidence level, keeping the consistency with that in Fig.~\ref{IKM_1}.
    Note that the lower semantic confidence level is generally inapplicable in practice, thus it may become tricky to consider in the IKM-based SC-Net how to strike a good balance between the preset risk level and the desired STM.
	As for the effect of $\tau$, as expected that the higher knowledge matching degree still enables the better STM performance as shown in Fig.~\ref{IKM_2}.
	Furthermore, we can always see a higher STM performance obtained by our solution when compared with the two baselines, and a slow growth trend of STM is also observed in all solutions as the number of BSs increases, which can be credit to more available bandwidth at MUs.
	
	Finally, if we laterally compare the results in both PKM-based and IKM-based SC-Nets, it can be concluded that when the knowledge matching state of MUs changes from PKM to IKM, the penalty of STM performance is inevitable.
	Taking Fig.~\ref{PKM_MU} and Fig.~\ref{IKM_1} as examples, in the same simulation settings, we observe an STM result around $42$ kmsg/s by our solution in the PKM case, while only $25$ kmsg/s STM is obtained in the IKM case.
	Due to the mismatching of partial KBs, the message generation and interpretation ability of IKM-based semantic coding models cannot be fully leveraged, which can incur a certain degree of semantic ambiguity compared with that in the PKM case.
	Therefore, it is of paramount importance to guarantee adequate knowledge matching degrees in SemCom to render a better network performance in the SC-Net.
	
	\section{Conclusions}
	This paper conducted a systematic study on SemCom from a networking perspective.
	Specifically, two typical scenarios of PKM-based and IKM-based SC-Nets were first identified, by which we presented their respective semantic channel models in combination with the existing works related to semantic information theory.
	After that, the concept of B2M transformation along with the new network performance metric STM were introduced in the two SC-Net scenarios, respectively.
	Then we formulated the joint optimization problem of UA and BA for each SC-Net scenario, followed by the corresponding solution proposed with the aim of STM maximization.
	Simulation results of both SC-Net scenarios demonstrated that our proposed solutions can always outperform two traditional benchmarks in terms of STM. 
	
	This paper promises to be a seminal work providing valuable inspirations for several follow-up research topics on wireless SemCom.
	For instance, it turns out that KB matching should be a crucial factor to affect the quality of SemCom service provisioning and the STM performance of SC-Net, therefore, a pressing need of effective KB matching algorithms inevitably arises in future research venues.
	Besides, this work provides foundations to properly generalize resource management solutions to other complicated SC-Net scenarios, e.g., PKM-IKM-coexistence networks.
	Moreover, some insights obtained from this work should inspire new resource management strategies in alignment with different new SemCom-relevant objectives, such as the accuracy of message interpretation, latency of end-to-end links, and user fairness in a semantical sense.
	
	\bibliographystyle{IEEEtran}
	\bibliography{main}
	
	\begin{IEEEbiography}[{\includegraphics[width=1in, height=1.25in, clip, keepaspectratio]{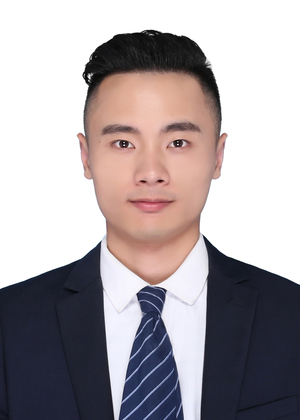}}]{Le Xia} (Graduate Student Member, IEEE)
	 received the B.Eng. degree in communication engineering and the M.Eng. degree in electronics and communication engineering from the University of Electronic Science and Technology of China (UESTC) in 2017 and 2020, respectively. He is currently pursuing his Ph.D. degree with James Watt School of Engineering, the University of Glasgow, United Kingdom. His research interests include semantic communications, intelligent vehicular networks, and resource management in next-generation wireless networks.
	\end{IEEEbiography}

	\begin{IEEEbiography}[{\includegraphics[width=1in, height=1.25in, clip, keepaspectratio]{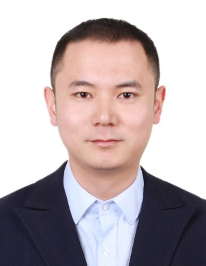}}]{Yao Sun} (Senior Member, IEEE)
	is currently a Lecturer with James Watt School of Engineering, the University of Glasgow, Glasgow, UK. Dr Sun has won the IEEE Communication Society of TAOS Best Paper Award in 2019 ICC, IEEE IoT Journal Best Paper Award 2022 and Best Paper Award in 22nd ICCT. His research interests include intelligent wireless networking, semantic communications, blockchain system, and resource management in next generation mobile networks. Dr. Sun is a senior member of IEEE.
	\end{IEEEbiography}

	\begin{IEEEbiography}[{\includegraphics[width=1in, height=1.25in, clip, keepaspectratio]{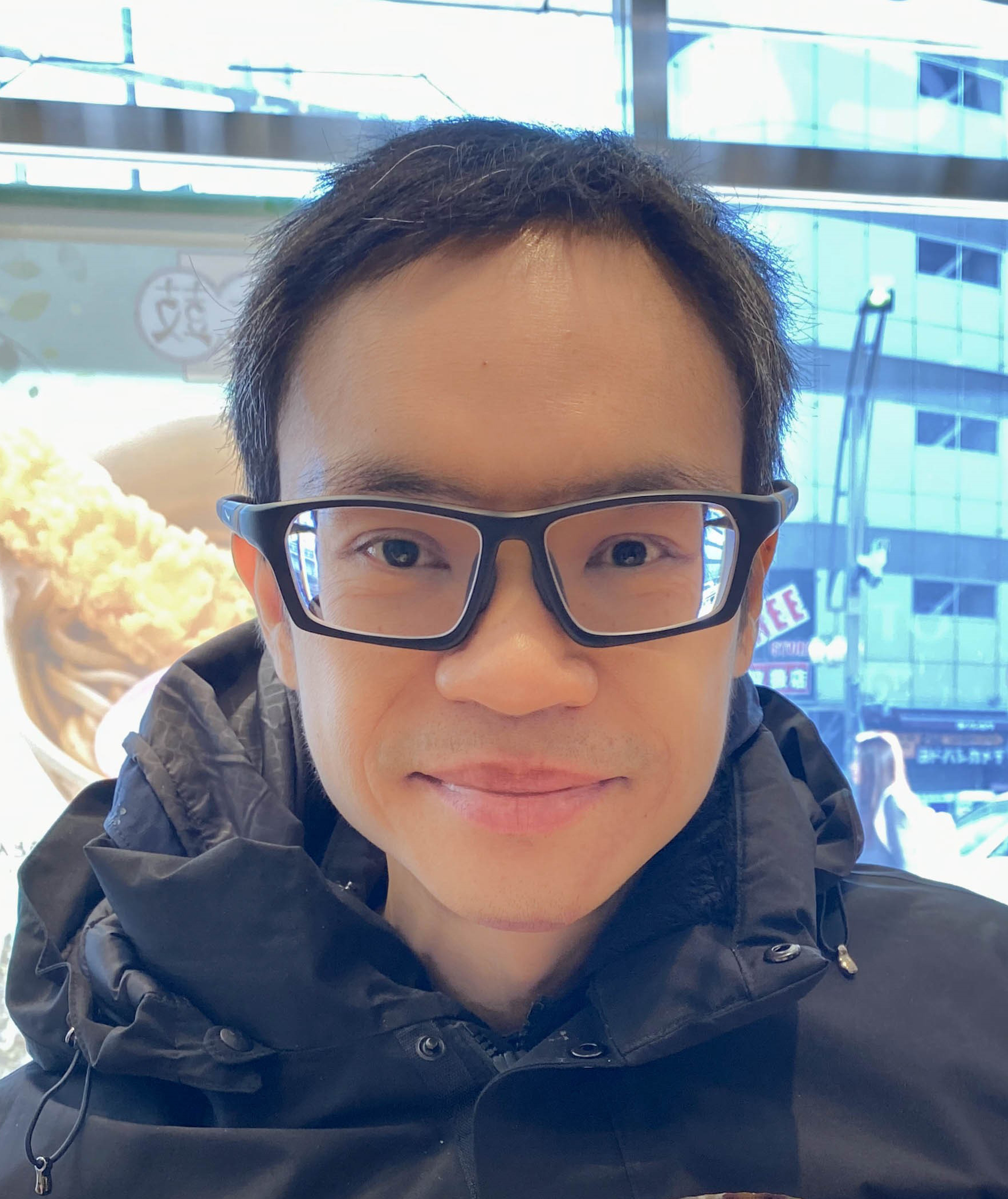}}]{Dusit Niyato} (M'09-SM'15-F'17, IEEE)
	  is a professor in the School of Computer Science and Engineering, at Nanyang Technological University, Singapore. He received B.Eng. from King Mongkut's Institute of Technology Ladkrabang (KMITL), Thailand in 1999 and Ph.D. in Electrical and Computer Engineering from the University of Manitoba, Canada in 2008. His research interests are in the areas of sustainability, edge intelligence, decentralized machine learning, and incentive mechanism design.
	\end{IEEEbiography}

	\begin{IEEEbiography}[{\includegraphics[width=1in, height=1.25in, clip, keepaspectratio]{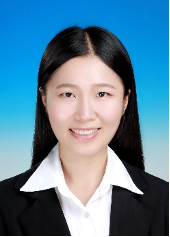}}]{Xiaoqian Li} (Member, IEEE)
	  received the B. Eng. and M. Eng. degrees in communication engineering from the University of Electronic Science and Technology of China (UESTC), Chengdu, China, in 2013 and 2016, respectively, and the Ph.D. degree in communication engineering from University of Hong Kong in 2020. She is currently an associate professor in the UESTC. Her research interests include next-generation Internet, mobile edge computing, and mobile crowd sensing.
	\end{IEEEbiography}

	\begin{IEEEbiography}[{\includegraphics[width=1in, height=1.25in, clip, keepaspectratio]{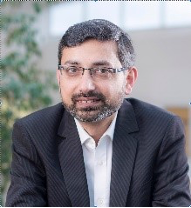}}]{Muhammad Ali Imran} (Fellow, IEEE) 
	received his M.Sc. (Distinction) and Ph.D. degrees from Imperial College London, UK, in 2002 and 2007, respectively. He is a Professor in Communication Systems in the University of Glasgow. He has a global collaborative research network spanning both academia and key industrial players in the field of wireless communications. He has supervised 50+ successful PhD graduates and published over 600 peer-reviewed research papers including more than 100 IEEE Transaction papers. Prof. Imran is a Fellow of IEEE.
	\end{IEEEbiography}
\end{document}